\documentclass[prb,amsmath,amssymb,superscriptaddress]{revtex4}
\usepackage{amsmath}
\usepackage{amssymb}
\usepackage{multirow}
\usepackage{abraces}
\usepackage{mathtools}
\usepackage{ltablex,longtable,booktabs,float}
\usepackage{sidecap}
\usepackage{yhmath}
\usepackage{dsfont}

\newcommand{\Tr}{{\mathop{\textrm{Tr}}}}

\newcommand\reverse[1]{\widetilde{#1}}
\newcommand\gradeinverse[1]{\wideparen{#1}}
\newcommand\cliffordconjugate[1]{\widetilde{\wideparen{#1}}}

\def\e#1{\mathbf{e}_{#1}} 

\newcommand{\ii}{\mathrm{i}} 
\newcommand{\jj}{\mathrm{j}}
\newcommand{\kk}{\mathrm{k}}

\newcommand{\bbR}{\ensuremath{\mathbb{R}}}

\newcommand{\bbC}{\ensuremath{\mathbb{C}}}
\newcommand{\bbH}{\ensuremath{\mathbb{H}}}

\newcommand{\bbZ}{\ensuremath{\mathbb{Z}}}
\newcommand{\Z}{\ensuremath{\mathds{Z}}}

\newcommand{\cl}[2]{\ensuremath{\mathit{Cl}_{#1,#2}}}

\newcommand{\cI}{\ensuremath{\mathcal{I}}}

\newcommand{\cR}{\ensuremath{\mathcal{R}}}

\newcommand{\cV}{\ensuremath{\mathcal{V}}}

\DeclarePairedDelimiterX{\norm}[1]{\lVert}{\rVert}{#1}

\begin{document}

\title{Clifford geometric algebra: Real and complex spinor data tables}
\author{A.\ Acus}\email{arturas.acus@tfai.vu.lt}
\affiliation{Institute of Theoretical Physics and Astronomy, Vilnius University,
Saul\.{e}tekio 3, LT-10222 Vilnius, Lithuania}

\author{A.~Dargys}\email{dargys@ftmc.lt}
\affiliation{Center of Physical Sciences and Technology,\\
 Semiconductor Physics Institute,  Saul\.{e}tekio 3, LT-10222 Vilnius, Lithuania}

\begin{abstract}
The modern algebra concepts are used to construct tables of algebraic spinors related to Clifford algebra multivectors with real and complex
coefficients. The following data computed  by \textit{Mathematica} are presented in form of tables for individual Clifford geometric algebras:
1.~Initial idempotent; 2.~Two-sided ideal; 3.~Left ideal basis (otherwise projector, or spinor basis); 4.~Matrix representations (reps) for basis vectors in Clifford algebras in spinor basis; 5.~General spinor; 6.~Spinor in matrix form; 7.~Squared hermitian norm of the spinor. Earlier in 1998, only the first four
items computed by \textit{Maple} were published by R.~Ab\l amowicz~\cite{Ablamowicz98}.
\end{abstract}

\maketitle

\section{Terms and definitions}\label{definitions}

Below, the definitions of algebraic concepts on which the computation is based are reminded. The starting structures are the ring $\cR$
and ideal $\cI$ the properties of which  are summarized below. The small circle "$\circ$" denotes a non-commutative
multiplication. In geometric algebra (GA) \cl{p}{q}, where $(p,q)$ and $n=p+q$ define  signature and dimension of the vector space,
respectively, the circle represents the geometric product (GP) that usually is  omitted in formulas.

\vspace{3mm} \textit{Division ring}~$\cR$ consists of a set of elements $A,B,C\dots\in\cR$ and binary
operations "+" (summation) and "$\circ$" (multiplication) defined on $\cR$. The properties of $\cR$ are summarized in Table~\ref{table1}. The
division ring $\cR$ has two units: "0" for addition and "1" for multiplication. The division rings commonly used are $\bbR$, $\bbC$
(commutative) and $\bbH$ (noncommutative). Commutative division rings are called fields. For commutative fields the inequality $A\circ B\ne
B\circ A$ in the table turns into equality.  All other properties remain unchanged. Quaternion division ring sometimes is referred to as
non-commutative field, though this is not overwhelmingly accepted. In GA, therefore,  $\bbR,\bbC$ and $\bbH$, collectively are called
fields, or just $\mathbb{K}$ field. In the ideal theory the description of non-commutative GA rings produces a new
GA structure of multivectors (MVs)  named the spinor.

\begin{table}
\label{table1}
\[\mspace{-50mu} \begin{array}{lll}\hline
\textrm{Name} &\mspace{-10mu}\textrm{Addition} &\mspace{-30mu}\textrm{Multiplication} \\
\hline
\textrm{Commutativity} &\mspace{-10mu} A+B=B+A&\mspace{-30mu}A\circ B\ne B\circ A \\
\textrm{Associativity} &\mspace{-10mu}(A+B)+C=A+(B+C)&\mspace{-30mu}(A\circ B)\circ C=A\circ (B\circ C) \\
\textrm{Distributivity} &\mspace{-10mu} A\circ(B+C)=(A\circ B)+(A\circ C)&\mspace{-30mu}\\
\textrm{Identity} &\mspace{-10mu}A+0=A=0+A &\mspace{-30mu} A\circ 1=A=1\circ A\\
\textrm{Inverse} &\mspace{-10mu}A+(-A)=0=(-A)+A &\mspace{-30mu} AA^{-1}=1=A^{-1}A\\
\hline
\end{array}
\]\caption{Properties of noncommutative division ring $\cR\ni A,B,C$} \end{table}

\textit{Ring algebra.} If $A\in\cR$ and requirement $AA^{-1}=A^{-1}A=1$ is released, then the set of properties (axioms) listed in the
Table~\ref{table1} describes a noncommutative ring $\cR$. The property of distributivity in this case splits into the left distributivity,
$A\circ(B+C)=(A\circ B)+(A\circ C)$, and the right distributivity, $(B+C)\circ A=(B\circ A)+(C\circ A)$. The simplest example of noncommutative
ring is a set of  $n\times n$ real matrices. The algebra satisfies the same set of axioms as a ring. Formally, the algebra is a vector space
$\cV$ over a field $\mathbb{K}$ with a multiplication which turns it into a ring $\cR$. The algebra also may have an additional
definition, in particular, describing how to multiply an algebra element by a scalar (or external ring of scalars). An external ring of
scalars in many cases, for example, in Clifford algebra, also may be internal, i.e., the scalars may belong to elements of algebra. A
\textit{division ring} or division algebra is a ring in which every nonzero element has a multiplicative inverse, but multiplication is not
necessarily commutative. A commutative division algebra is also called a field.

The \textit{left (right) ideal} $\cI$ is a subset of the ring $\cR$ which has an additional property: if $A\in\cI$ is an ideal element and
$B\in\cR$ is any ring element then the noncommutative product $B\circ A\in\cI$, (respectively $A\circ B\in\cI$), i.e. belongs to left (right)
ideal too. For example, the set of even integers is an ideal in the ring of all integers $\bbZ$, since any integer multiplied by even number is
even. Also a matrix with one nonzero column (and other elements filled with zeroes) is left ideal element, since multiplication of it by arbitrary matrix from the left again gives a matrix with single column. In case of commutative  multiplication the left and right ideals coincide and the prefix left/right is omitted. The ideals allow to obtain spinor basis and construct the spinor.

\textit{Idempotents.} Basic definitions and some basic facts about the idempotents and their relation to the ideals in GA are presented in
textbook~\cite{Lounesto1997}. An element $P$ of a ring $\cR\in\cl{p}{q}$ is called an idempotent
 if $P^2=P$. For example, $\frac12 (1+\e{3})$ is an idempotent in $\cl{3}{0}$ algebra.
Two idempotents $P_1$ and $P_2$ are \textit{annihilating} if $P_1P_2=P_2P_1=0$. For example, $P_1=\frac12 (1+\e{1})$ and $P_2=\frac12 (1-\e{1})$
are annihilating idempotents\footnote{In tables~\ref{spt1}-\ref{spt6b}  and \ref{sptc1}-\ref{sptc5} we have used a shorthand notation:
$P_1=\frac12(1+\e{1})$, $P_2=(-)\equiv\frac12(1-\e{1})$.} in $\cl{3}{0}$. The idempotents $P_i$, $i=1,2,\dots,m$, in a ring $\cR$ are called
\textit{mutually annihilating} if and only if they are annihilating in all pairs, i.e., $P_iP_j=0$ for all $i\ne j$. An idempotent is
\textit{primitive}, for example $\frac12(1+\e{1})$ is the primitive idempotent, if it is not a sum of two nonzero annihilating
idempotents. An idempotent $P_2$ is \textit{minimal} if it is a minimal element in the set of all nonzero (not annihilating) idempotents $P_i$  with the order relation $P_2\le P_1$  and  iff $P_1P_2=P_2P_1=P_2$. 

\textit{Construction of full set of mutually annihilating idempotents.}
If $\{\e{T_1},\e{T_2},\dots,\e{T_k}\}$, where $T_k$ denotes a multi-index, is a set of commuting GA
elements that square to $1$, i.e. $\e{T_i}^2=1$, then allowing the signs of
the elements $\e{T_i}$ in the product
\[P_1=\tfrac{1}{2}(1+\e{T_1})\tfrac{1}{2}(1+\e{T_2})\cdots \tfrac{1}{2}(1+\e{T_k})\]
to vary independently, i.e.,
\[\begin{matrix}
P_2=\tfrac{1}{2}(1-\e{T_1})\tfrac{1}{2}(1+\e{T_2})\cdots
\tfrac{1}{2}(1+\e{T_k}),\\
P_3=\tfrac{1}{2}(1+\e{T_1})\tfrac{1}{2}(1-\e{T_2})\cdots
\tfrac{1}{2}(1+\e{T_k}),\\
\hdotsfor[2]{1}\\
P_{2^k}=\tfrac{1}{2}(1-\e{T_1})\tfrac{1}{2}(1-\e{T_2})\cdots
\tfrac{1}{2}(1-\e{T_k}),
\end{matrix}
\]
one obtains $2^k$ idempotents which are mutually annihilating and
sum up to unity, $P_1+P_2+\cdots +P_{2^k}=1$. Then it is said that
$\cl{p}{q}$ algebra decomposes into a direct sum of $2^k$
ideals. By construction, each of the ideals (see definition below) has
dimension $2^{n-k}$. For real algebras the above maximal product
of non-annihilating and commuting idempotents, which square to
one, yields a set of minimal left ideals 
$\cI(P_i)=\{\cl{p}{q}\}P_i$ at fixed $1\le i\le n-k$.

\textit{Number of primitive idempotents.}
In  real GAs  the construction of primitive idempotent requires $(q-r_{q-p})$ commuting blades which square to $+1$. Here $r_i$
denotes the Radon-Hurwitz number, which can be computed as a
series coefficients of generating function $\frac{x (1+ x + x^3 +
x^7)}{(1 - x)(1 - x^8)}$. First few coefficients are 
$r_0=0$, $r_1=1$, $r_2=r_3=2$, $r_4=r_5=r_6=r_7=3$. Higher numbers
can be handled by recursive relation $r_{i+8}=r_i+4$. The Radon-Hurwitz numbers (including negative indices) are listed in Table~\ref{Hurwitz}.
Negative Radon-Hurwitz numbers are calculated by $r_{-i}=1-i+r_{i-2}$ ($i>0$) where $r_{-1}=-1$.
\begin{table}
\label{Hurwitz}
\[\mspace{-50mu} \begin{array}{c|ccccccccccccccccc}
\hline\\
 i&-8&-7&-6&-5&-4&-3&-2&-1&0&1&2&3&4&5&6&7&8\\
 r_i&-4&-3&-2&-2&-1&-1&-1&-1&0&1&2&2&3&3&3&3&4\\ \hline
\end{array}
\]\caption{Radon-Hurwitz numbers $r_i$}
\end{table}
In the canonical (standard) basis of $\cl{p}{q}$ there are always $k=q-r_{q-p}$
commuting elements $\e{T_i}$, with property $\e{T_i}^2=1$ that
generate the idempotents $\tfrac{1}{2}(1+\e{T_i})$. The products of $k$ such
idempotents is a primitive idempotent
$P=\frac{1}{2}(1+\e{T_1})\cdots\frac{1}{2}(1+\e{T_k})$.
The statement does not apply to complex Clifford algebras.

\textit{Left and right ideals.} A subring $\cI\subset\cl{p}{q}$ is a \textit{left ideal} if $ax\in \cI$ for all $a\in\cl{p}{q}$ and $x\in \cI$.
Similarly, for the \textit{right ideals} we have $xa\in \cI$ for all $a\in\cl{p}{q}$ and $x\in \cI$.  Here we shall be interested in left ideals only. 

\textit{Minimal left ideal} $\cI$. In geometric algebra \cl{p}{q} where $p+q=n$, any minimal left ideal is of the form $\cI=\{\cl{e}{..}\}P_i$
for some primitive idempotent $P_i$ and for  a list of generators $\{\cl{e}{..}\}=\{\e{1},\e{2},\dots,\e{n}\}$. An ideal generated by idempotent
$P_i$ will be denoted as $\cI(P_i)$. The list $\{\cl{e}{..}\}P_i$ of minimal ideal does not contain other ideals except trivial ones
(itself and zero ideal).

\textit{Two-sided (bilateral) ideals $\mathbb{K}$.}  The two-sided ideal $K_i$ generated by primitive idempotent $P_i$ is defined by
$K_i=P_i\cI=P_i\{\cl{p}{q}\}P_i$, where $\{\cl{p}{q}\}$ is a list of basis elements of \cl{p}{q}. According to  periodicity table, in Clifford
geometric algebra two-sided ideals represent real numbers $\bbR$, complex numbers $\bbC$ or quaternions $\bbH$.
Thus, the two-sided ideal $K_i$ for real $\cl{p}{q}$ algebras contains $k=1$, $2$, or $4$ different elements that play respectively the role of $\bbR$,
$\bbC$ or $\bbH$ division rings (fields), correspondingly.

\textit{Spinor basis (generators, or projectors) $S$.}  The basis can be obtained from single sided ideal, $S=\{\cl{p,q}\}P_i$ once elements that belong to two sided ideal are replaced by division ring elements.
The number of the spinor basis elements is equal to dimension of matrix representation (rep) of GA in Bott's periodicity
table~\cite{Lounesto1997}.

\textit{Spinor} $\Psi$. The spinor $\Psi\in \cI(P_i)$, can be expressed as a sum of all spinor generators $S_i$ (multiplied by coefficients
$s_k\in\bbR$ for real idempotents and $s_k\in\bbC$ for complex idempotents). Hat over spinor $\hat{\Psi}$ denotes matrix rep of a spinor and $\norm{\Psi}^2$ is the spinor norm.

\section{Ordering of basis elements in GA}
Tables~\ref{spt1}-\ref{sptc5} lists the idempotent factors in the order of growing numbers in a blade index, exactly  as letters are listed in the alphabet. Sometimes different ordering of basis
elements is needed. In GA four main orderings of basis elements
are defined: lexicographic (Lex), inverse lexicographic
(InvLex), reverse lexicographic (RevLex), inverse reverse
lexicographic (InvRevLex). These orderings may be preceded by
two possible corrections "Deg" (higher degree blades first) or
"InvDeg" (higher degree blades last), so all in all we have  $4\times 3=12$ possible orderings of GA basis elements. We will denote the
additional orderings as Deg[InvLex], Deg[RevLex],
Deg[InvRevLex], InvDeg[Lex], InvDeg[InvLex],
InvDeg[RevLex] and InvDeg[InvRevLex]. To distinguish the named
orderings an algebra of dimension $n\ge4$ is required. Given
two sets of indices (treated as components of abstract vector in
the index space) $\alpha=\{\alpha_1,\alpha_2\dots\}$ and
$\beta=\{\beta_1,\beta_2\dots\}$ with their lengths (which encode
the grade of the element) denoted as $|\alpha|$ and $|\beta|$ the
mentioned orderings are defined in the following way:
\begin{itemize}
\setlength{\parskip}{0pt} \setlength{\itemsep}{0pt plus 1pt}
\setlength{\itemindent}{.3in}
 \item[\footnotesize\bf Lex]{ordering, $\alpha >_\mathrm{Lex} \beta$, if in the vector
difference $\alpha -\beta \in \Z^n$ the {\it left most} nonzero
entry is {\it positive} (for index set with less elements zeroes
are inserted in the place of absent index, see
    subsection~\ref{ExampleOrdering} below). An
example (\textit{Ex.}) of Lex ordering is
\newline \textit{Ex.} $\{\e{1234}, \e{123}, \e{124}, \e{12}, \e{134}, \e{13},
\e{14}, \e{1}, \e{234}, \e{23}, \e{24}, \e{2}, \e{34}, \e{3},
\e{4}, 1\}$.}

  \item[\footnotesize\bf InvLex]{ordering, $\alpha >_\mathrm{InvLex} \beta$, if in
  $\alpha -\beta \in \Z^n$ the {\it right most} nonzero entry is {\it
  positive}.\\
  \textit{Ex.}: $\{\e{1234},\e{234},\e{134},\e{34},\e{124},\e{24},\e{14},\e{4},\e{123},\e{23},\e{13},\e{3},\e{12},
   \e{2}, e{1},1\}$.}

  \item[\footnotesize\bf RevLex]{ordering, $\alpha >_\mathrm{RevLex} \beta$, if in
  $\alpha -\beta \in \Z^n$ the {\it right most} nonzero entry is {\it
  negative}.\\
  \textit{Ex.}: $\{1,\e{1},\e{2},\e{12},\e{3},\e{13},\e{23},\e{123},\e{4},\e{14},\e{24},\e{124},\e{34},\e{134}, \e{234},\e{1234}\}$.}

  \item[\footnotesize\bf InvRevLex] {ordering, $\alpha >_\mathrm{InvRevLex}
  \beta$,  if in  $\alpha -\beta \in \Z^n$ the {\it left most} nonzero entry is {\it
  negative}.\\
  \textit{Ex.}: $\{1,\e{4},\e{3},\e{34},\e{2},\e{24},\e{23},\e{234},\e{1},\e{14},\e{13},\e{134},\e{12},\e{124},
  \e{123},\e{1234}\}$.}

  \item[{\footnotesize\bf Deg[..]}]

  \item[{\footnotesize\bf Lex}] {ordering, $\alpha >_\mathrm{Deg[Lex]} \beta$,
  if  $|\alpha|>  |\beta|$ and Lex when $|\alpha| = |\beta|$.\\
  \textit{Ex.}: $\{\e{1234},\e{123},\e{124},\e{134},\e{234},\e{12},\e{13},\e{14},\e{23},\e{24},\e{34},\e{1},\e{2},\e{3},\e{4},1\}$.}

  \item[{\footnotesize\bf InvLex}] {ordering, $\alpha >_\mathrm{Deg[InvLex]} \beta$,
  if $|\alpha|> |\beta|$ and InvLex when $|\alpha| = |\beta|$.\\
  \textit{Ex.}: $\{\e{1234},\e{234},\e{134},\e{124},\e{123},\e{34},\e{24},\e{14},\e{23},\e{13},\e{12},\e{4},\e{3},\e{2},\e{1},1\}$.}

  \item[{\footnotesize\bf RevLex}] {ordering, $\alpha >_\mathrm{Deg[RevLex]} \beta$,
  if $|\alpha|> |\beta|$ and RevLex when $|\alpha| = |\beta|$.\\
  \textit{Ex.}: $\{\e{1234},\e{123},\e{124},\e{134},\e{234},\e{12},\e{13},\e{23},\e{14},\e{24},\e{34},\e{1},\e{2},\e{3},\e{4},1\}$.}

  \item[{\footnotesize\bf InvRevLex}]
  {ordering, $\alpha >_\mathrm{Deg[InvRevLex]} \beta$, if $|\alpha| > |\beta|$
  and InvRevLex when $|\alpha| = |\beta|$.\\
 \textit{Ex.}: $\{\e{1234},\e{234},\e{134},\e{124},\e{123},\e{34},\e{24},\e{23},\e{14},\e{13},\e{12},\e{4},\e{3},\e{2},
   \e{1},1\}$.}
  \item[{\footnotesize\bf InvDeg[..]}]
 \item[{\footnotesize\bf Lex}] {ordering, $\alpha >_\mathrm{InvDeg[Lex]} \beta$, if  $|\alpha| < |\beta|$ and Lex when $|\alpha| = |\beta|$.\\
 \textit{Ex.}: $\{1,\e{1},\e{2},\e{3},\e{4},\e{12},\e{13},\e{14},\e{23},\e{24},\e{34},\e{123},\e{124},\e{134},\e{234},\e{1234}\}$}

  \item[{\footnotesize\bf InvLex}] {ordering, $\alpha >_\mathrm{InvDeg[InvLex]} \beta$, if $|\alpha|<  |\beta|$ and InvLex when $|\alpha| = |\beta|$.\\
  \textit{Ex.}: $\{1,\e{4},\e{3},\e{2},\e{1},\e{34},\e{24},\e{14},\e{23},\e{13},\e{12},\e{234},\e{134},\e{124},\e{123},\e{1234}\}$}

  \item[{\footnotesize\bf RevLex}] {ordering, $\alpha >_\mathrm{InvDeg[RevLex]} \beta$, if $|\alpha| <|\beta|$ and RevLex when $|\alpha| = |\beta|$.\\
  \textit{Ex.}: $\{1,\e{1},\e{2},\e{3},\e{4},\e{12},\e{13},\e{23},\e{14},\e{24},\e{34},\e{123},\e{124},\e{134},\e{234},\e{1234}\}$}

  \item[{\footnotesize\bf InvRevLex}] {ordering, $\alpha >_\mathrm{InvDeg[InvRevLex]} \beta$,
  if $|\alpha| < |\beta|$ and InvRevLex when $|\alpha| =
  |\beta|$.\\
  \textit{Ex.}: $\{1,\e{4},\e{3},\e{2},\e{1},\e{34},\e{24},\e{23},\e{14},\e{13},\e{12},\e{234},\e{134},\e{124},
  \e{123},  \e{1234}\}$}
\end{itemize}
Orderings RevLex, InvRevLex and the default ordering InvDeg[Lex] are not admissible as Gr\"obner basis monomial orderings.
GA basis element ordering plays an important role in automatic computation of spinors. In particular if we want to construct a spinor, represented by a first matrix column we have to sort ideal basis elements in a specific order (see next section).

\subsection{Example of blades $\e{12}$ and $\e{134}$ ordering\label{ExampleOrdering}}
 Given two basis elements $\e{12}$ and $\e{134}$
let's  determine their position in all orderings. After insertion
of zeroes we have $\{1,2,0,0\}-\{1,0,3,4\}=\{0,2,-3,-4\}$. The
first nonzero element in the {\it left most} position is~$2$. It
is positive, therefore in Lex $\e{12}>_\mathrm{Lex}\e{134}$.
Similarly, for the same reason in InvLex order
$\e{12}<_\mathrm{InvLex}\e{134}$, because the {\it right most}
element $-4$ is negative. For the same reason in RevLex
$\e{12}>_\mathrm{RevLex}\e{134}$. Finally, for InvRevLex the left
most nonzero element is $2$, therefore
$\e{12}<_\mathrm{InvRevLex}\e{134}$. It can be checked that the examples above incorporate both elements in the correct relative positions. Of course, in Deg[Lex], Deg[InvLex],
Deg[RevLex] and Deg[InvRevLex] one has
$\e{12}<_\mathrm{Deg[~]}\e{134}$ and in InvDeg[Lex],
InvDeg[InvLex], InvDeg[RevLex] and InvDeg[InvRevLex] one has
$\e{12}>_\mathrm{InvDeg[~]}\e{134}$.

\section{Real GA spinors as minimal left ideals\label{realGAspinors}}
The procedure to compute algebraic spinors and their matrix representations (reps) are described in detail in
papers~\cite{Ablamowicz98,Ablamowicz2005}, that are adapted to {\it Mathematica} and further improved by adding automatic ordering of
ideal basis elements~\cite{AcusDargys2024}, spinors and their norms.

An algebraic spinor computation procedure is initiated by choosing a primitive idempotent. After a primitive idempotent of  respective algebra
is chosen one consecutively calculates the left ideals, two-sided ideals, left ideal basis (generators), respective matrix reps and finally the
algebraic spinors. The obtained data  are summarized  in the Tables~\ref{spt1}-\ref{spt6b} for real idempotents/spinors and
Tables~\ref{sptc1}-\ref{sptc5} for complex idempotents/spinors. The items  in the tables are enumerated from $1$ to $7$ and carry the following
information:
\begin{itemize}
\item[1.] defines a list of \textit{mutually annihilating idempotents} $P_i\in\cl{p}{q}$ with factors sorted in InvDeg(Lex)
order. The idempotents that
differ by signs are singled out by plus/minus signs. For example, $P_1=\tfrac12(1+ \e{1})\equiv(+)$ and $P_2=\tfrac12(1-\e{1})\equiv(-)$, or
$P_1=\tfrac14(1+ \e{1})(1+\e{23})\equiv(+,+)$ and   $P_2=\tfrac14(1 + \e{1})(1-\e{23})\equiv(+,-)$.
\item[2.] gives a \textit{list of elements of two-sided ideal} $K_1=\{K_1(1)\}$ , or $K_1=\{K_1(2)\}$, or  $K_1= \{K_1(4)\}$  generated by $K_1=P_1\{\cl{p}{q}\}P_1$.
  The number of elements, denoted as (1),(2) or (4), in the list of two-sided ideal determines a type of matrix rep: $1\leftrightarrow \bbR$, $2\leftrightarrow \bbC$, and $4\leftrightarrow \bbH$, i.e. if $K_1$ contains single element (elements that differ by sign are considered same element) which squares to $+1$ then the matrix rep entries are real numbers (matrix representation is real). When there are two different elements (one squares to $+1$, other to $-1$ and both commute), then the matrix representation entries are complex numbers (matrix representation is complex). Lastly, if there are four different non-zero elements then the matrix rep entries are quaternions and respective four elements $K(4)$ have to be replaced by quaternion units $1,q_1,q_2$, and $q_3$ by matching their properties.
\item[3.] provides \textit{spinor basis} $S_1=\{S_1(1),S_1(2),\ldots\}P_1$ for the left ideal spinors $\cI(P_1)$. The individual elements
$S_1(i)$ are called the \textit{projections} of the ideal $ \cI(P_1)$, or just \textit{generators}. In quantum mechanics they represent
\textit{spinor basis} of a respective geometric algebra. The basis is sorted in RevLex order that ensures that the
matrix rep of spinor will contain only the leftmost nonzero matrix column. The length of the ideal basis determines the dimension of the matrix
rep (for simple algebras) and dimension of the block for the block-diagonal matrix (for semisimple algebras). In the latter case the remaining
    half of the ideal basis can be obtained by grade inversion and should be sorted in InvLex order to ensure proper place of the column of matrix representation of second block.
\item[4.] gives the matrix rep of basis vectors $\e{k}$ in the minimal left ideal basis $S_1$. The matrix type and
    dimension is in agreement with the 8-periodicity table~\cite{Lounesto1997}. The symbol
$E_{ij}$ here denotes  matrix with a single entry equal to $1$ at the intersection of the $i$-th row and $j$-th column. For example,
 $E_{11}-E_{22}=\left[\begin{smallmatrix}1&0\\
0&-1\\ \end{smallmatrix}\right]$, or $q(-E_{12}+E_{21})=\left[\begin{smallmatrix}0&-q\\
q&0\\ \end{smallmatrix}\right]$, where $q$ is the quaternion $q=q_0+a q_1+b q_2 +c q_3$ and $q_1^2=q_2^2=q_3^2=-1$. The first of matrices is
diagonal what indicates that the quantization axis of the spinor is parallel to GA basis vector~$\e{1}$ (this property is determined by choice
of an idempotent  in item 1.).

\item[5.] lists general spinor $\Psi$ expanded in GA minimal left ideal basis, where $s_i$~are the scalar expansion coefficients. In
tables the spinor expressions  for semisimple algebras  includes two idempotents that are related by grade inversion (denoted by $\gradeinverse{P_i}$).
\item[6.] gives the matrix rep $\hat{\Psi}$ of the GA general spinor~$\Psi$.
\item[7.] expresses the square of the spinor norm $\norm{\Psi}^2$,
  which is a positive scalar. The spinor hermitian norm can be written universally (in algebra independent way), as
    $\norm{\Psi}^2=\langle\Psi^\dagger\Psi\rangle$, where $\Psi^\dagger=\sum_T c_T^*\e{T}^{-1}$ is the spinor in the reciprocal basis, an asterisk $*$ denotes complex conjugation of scalar coefficient and $T$ is a multi-index, i.e. the sum is over all ordered orthonormal basis. The universal hermitian norm formula can be rewritten by algebra dependent combination of involutions presented in LHS of this item.
\end{itemize}
%


\section{Spinor data tables (real idempotents)\label{realGAspinorTable}}

The trivial primitive idempotents, $\{P_1=1\}$, in algebras $\cl{0}{1}$ and $\cl{0}{2}$  which, respectively, represent complex  $\bbC$ and
quaternion $\bbH$ numbers, suggest that the spinors in these algebras are division rings. In the tables below, $q_1\equiv\ii$,
$q_2\equiv\jj$ and $q_3\equiv\kk$, where squares of  quaternion units are $\ii^2=\jj^2=\kk^2=-1$. The arc over MV, for example
$\gradeinverse{\e{3}P_1}$, denotes grade inversion. The tilde $\reverse{\Psi}$ represents reversion. The combination of both is a Clifford conjugation $\cliffordconjugate{\Psi}$. 
\begin{flushleft}
\label{realSpinorTable}
\begin{table}[H]
\centering
\begin{tabular}{l|l>{$}l<{$}>{$}l<{$}}
  & Item number and  name & \cl{1}{0}[{}^2\bbR(1)] &\cl{0}{1}[\bbC(1)]
\\ \hline
1&\textrm{Idempotent, $P_i$}&
\{P_1=\tfrac12 (1 + \e{1}),P_2=(-)=\gradeinverse{P_1}\}
&
\begin{multlined}[t][0.1\columnwidth]
\{P_1=1 \}
\end{multlined}
\\
2&\textrm{Two-sided ideal}, $K_1$&
\{P_1\}
&
\begin{multlined}[t][0.1\columnwidth]
\{1,\e{1}\}
\end{multlined}
\\
3&\textrm{Projectors, (generators)}, $S_1$&
\{P_1\}\cup \{\gradeinverse{P_1}\}
&
\begin{multlined}[t][0.1\columnwidth]
\{P_1\}
\end{multlined}
\\
4&\textrm{Vector matrix rep, $\hat{e}_j$}&
\{E_{11} - E_{22}\}
&
\begin{multlined}[t][0.1\columnwidth]
\{\ii E_{11}\}
\end{multlined}
\\
5&\textrm{General spinor, $\Psi$}&\index{spinor}
\begin{multlined}[t][0.1\columnwidth]
  s_{1}P_1 + s_{2}\gradeinverse{P_1}
\end{multlined}
&
\begin{multlined}[t][0.1\columnwidth]
  s_1+s_2 \e{1}
\end{multlined}
\\
6&\textrm{Spinor in matrix form,
$\hat{\Psi}$}&
s_{1}E_{11} +s_{2} E_{22}
&
(s_1+s_2 \ii)E_{11}
\\
7&\textrm{Norm squared}, $\norm{\Psi}^2$&\index{norm}
\begin{multlined}[t][0.1\columnwidth]
\langle\Psi\Psi\rangle=\tfrac12(s_1^2+s_2^2)
\end{multlined}
&
\gradeinverse{\Psi}\Psi=s_1^2+s_2^2
\end{tabular}
\caption{Real spinor tables for GAs with vector space dimension $n=1$}\label{spt1}
\end{table}
\vskip 10pt

\begin{table}[H]
\centering
\begin{tabular}{l|>{$}l<{$}>{$}l<{$}>{$}l<{$}}
  &  \cl{2}{0}[\bbR(2)] & \cl{1}{1}[\bbR(2)] &\cl{0}{2}[\bbH]
  \index{algebra!\cl{2}{0}}\index{algebra!\cl{1}{1}}\index{algebra!\cl{0}{2}}
\\ \hline
1&
\begin{multlined}[t][0.1\columnwidth]
\{P_1=\tfrac12 (1 + \e{1}),\\[-3ex]P_2=(-)\}
\end{multlined}
&
\begin{multlined}[t][0.1\columnwidth]
\{P_1=\tfrac12 (1 + \e{1}),\\[-3ex]P_2=(-)\}
\end{multlined}
&
\begin{multlined}[t][0.1\columnwidth]
\{P_1=1 \}
\end{multlined}
\\
2&
\{P_1\}
&
\{P_1\}
&
\begin{multlined}[t][0.1\columnwidth]
\{1,\e{1}, \e{2}, \e{3}\}
\end{multlined}
\\
3&
\begin{multlined}[t][0.1\columnwidth]
  \{P_1,\e{2}P_1\}
\end{multlined}
&
\begin{multlined}[t][0.1\columnwidth]
 \{P_1,\e{2}P_1\}
\end{multlined}
&
\begin{multlined}[t][0.1\columnwidth]
\{P_1\}
\end{multlined}
\\
4&
\begin{multlined}[t][0.1\columnwidth]\{E_{11} - E_{22},\\[-3ex] E_{12} + E_{21}\}
\end{multlined}
&
\begin{multlined}[t][0.1\columnwidth]
\{E_{11} - E_{22},\\[-3ex] -E_{12} + E_{21}\}
\end{multlined}
&
\begin{multlined}[t][0.1\columnwidth]
\{q_1 E_{11}, q_2 E_{11}\}
\end{multlined}
\\
5&
\begin{multlined}[t][0.1\columnwidth]
\Psi=(s_{1}+s_{2}\e{2})P_1
\end{multlined}
&
\begin{multlined}[t][0.1\columnwidth]
\Psi=(s_{1}+s_{2}\e{2})P_1
\end{multlined}
&
\Psi=s_1+s_2 \e{1}+ s_3 \e{2}+s_4 \e{12}
\\
6&
s_{1}E_{11} +s_{2} E_{12}
&
s_{1}E_{11} +s_{2} E_{12}
&
(s_1+s_2 q_1+ s_3 q_2+s_4 q_3)E_{11}
\\
7&
\begin{multlined}[t][0.1\columnwidth]
\langle\reverse{\Psi}\Psi\rangle =\tfrac12(s_1^2+s_2^2)
\end{multlined}
&
\begin{multlined}[t][0.1\columnwidth]
\langle\e{1}\widetilde{\Psi}\e{1}\Psi\rangle
=\tfrac12(s_1^2+s_2^2)
\end{multlined}
&
\cliffordconjugate{\Psi}\Psi=s_1^2+s_2^2+s_3^2 +s_4^2
\end{tabular}
\caption{Real spinor table for  GAs with vector space dimension $n=2$}\label{spt2}
\end{table}
\vspace{5mm}

\begin{table}[H]
\centering
\begin{tabular}{l|>{$}l<{$}>{$}l<{$}>{$}l<{$}>{$}l<{$}}
& \cl{3}{0}[\bbC(2)]& \cl{2}{1}[{^2\bbR(2)}]
 \index{algebra!\cl{3}{0}}\index{algebra!\cl{2}{1}}\index{algebra!\cl{1}{2}}
\\ \hline
1&
\begin{multlined}[t][0.1\columnwidth]
  \{P_1=\tfrac12 (1 + \e{1}),P_2=(-)\}
\end{multlined}
&
\begin{multlined}[t][0.1\columnwidth]
\{P_1=\tfrac14 (1 + \e{1})(1 + \e{23}),
  P_2=(+,-),
  P_3=(-,+),
  P_4=(-,-)
\}
\end{multlined}
\\
2&
\begin{multlined}[t][0.1\columnwidth]
\{P_1 1\, P_1, P_1 \e{23} P_1\}
\end{multlined}
&
\{P_1 1\, P_1\}
\\
3&
\begin{multlined}[t][0.1\columnwidth]
\{P_1,\e{2} P_1\}
\end{multlined}
&
\begin{multlined}[t][0.1\columnwidth]
  \{P_1, \e{2} P_1\}\cup \{\gradeinverse{P_1}, \gradeinverse{\e{2} P_1}\}
\end{multlined}
\\
4&
\begin{multlined}[t][0.1\columnwidth]
  \{E_{11} - E_{22}, E_{12} + E_{21}, \ii (-E_{12} + E_{21})
\}
\end{multlined}
&
\begin{multlined}[t][0.1\columnwidth]\{E_{11} - E_{22} - E_{33} + E_{44},
  E_{12} + E_{21} - E_{34} - E_{43},\\[-3ex]
-E_{12} + E_{21} + E_{34} - E_{43}\}
\end{multlined}
\\
5&
\begin{multlined}[t][0.1\columnwidth]
  \Psi=(s_{1}  + s_{2}\e{23} + s_{3}\e{2} - s_{4}\e{3}) P_1
\end{multlined}
&
\begin{multlined}[t][0.1\columnwidth]
  \Psi=(s_{1} + s_{2}\e{2}) P_1 + (s_{3} - s_{4}\e{2}) \gradeinverse{P_1}
\end{multlined}
\\
6&
\begin{multlined}[t][0.1\columnwidth]
(s_{1} + \ii s_{2})E_{11} +(s_{3} - \ii s_{4}) E_{21}
\end{multlined}
&
\begin{multlined}[t][0.1\columnwidth]
s_{1}E_{11} +s_{2} E_{12}+s_{3}E_{33} +s_{4} E_{34}
\end{multlined}
\\
7&
\begin{multlined}[t][0.1\columnwidth]
\langle\reverse{\Psi}\Psi\rangle=\tfrac12(s_1^2+\cdots+s_4^2)
\end{multlined}
&
\begin{multlined}[t][0.1\columnwidth]
 - \langle
 \e{3}\cliffordconjugate{\Psi}\e{3}\Psi\rangle=\tfrac14(s_1^2+\cdots+s_4^2)
\end{multlined}
\\
\noalign{\bigskip}
& \cl{1}{2}[\bbC(2)]& \cl{0}{3}[\bbH(2)]
\\ \hline
1&
\begin{multlined}[t][0.1\columnwidth]
  \{P_1=\tfrac12 (1 + \e{1}), P_2=(-)\}
\end{multlined}
&
\begin{multlined}[t][0.1\columnwidth]
\{P_1=\tfrac12 (1 + \e{123}),
 P_2=\tfrac12(1 - \e{123})
 \}
\end{multlined}
\\
2&
\begin{multlined}[t][0.1\columnwidth]
  \{P_1 1\, P_1, P_1 \e{23} P_1\}
\end{multlined}
&
\begin{multlined}[t][0.1\columnwidth]
\{P_1 1 P_1, P_1 \e{1} P_1, P_1 \e{2} P_1, P_1 \e{3} P_1
 \}
\end{multlined}
\\
3&
\begin{multlined}[t][0.1\columnwidth]
\{P_1,
 \e{2}P_1\}
\end{multlined}
&
\begin{multlined}[t][0.1\columnwidth]
\{\e{3} P_1\}\cup \{\gradeinverse{\e{3} P_1}\}
\end{multlined}
\\
4&
\begin{multlined}[t][0.1\columnwidth]
  \{E_{11} - E_{22}, -E_{12} + E_{21}, -\ii E_{12} - \ii E_{21}
\}
\end{multlined}
&
\begin{multlined}[t][0.1\columnwidth]
\{-q_{1} E_{11} + q_{1}E_{22},
-q_{2}E_{11} + q_{2} E_{22},
-q_{3}E_{11} + q_{3} E_{22}
\}
\end{multlined}
\\
5&
\begin{multlined}[t][0.1\columnwidth]
 \Psi= (s_{1} + s_{2}\e{23} + s_{3}\e{2} + s_{4}\e{3}) P_1
\end{multlined}
&
\begin{multlined}[t][0.1\columnwidth]
  \Psi=(- s_{1} -s_{2}\e{1} + s_{3}\e{2} + s_{4}\e{3} )P_1
  +(s_{5} - s_{6}\e{1} + s_{7}\e{2} - s_{8}\e{3})\gradeinverse{P_1}
\end{multlined}
\\
6&
\begin{multlined}[t][0.1\columnwidth]
(s_{1} + \ii s_{2})E_{11} + (s_{3} - \ii s_{4})E_{21}
\end{multlined}
&
\begin{multlined}[t][0.1\columnwidth]
(-s_{1} +s_{2}q_1  - s_{3}q_2 - s_{4}q_{3})E_{11}
+(s_{5}- s_{6}q_{1} + s_{7}q_{2} - s_{8}q_{3})E_{22}
\end{multlined}
\\
7&
\begin{multlined}[t][0.1\columnwidth]
\langle\e{1}\reverse{\Psi}\e{1}\Psi\rangle=\tfrac12(s_1^2+\cdots+s_4^2)
\end{multlined}
&
\begin{multlined}[t][0.1\columnwidth]
\langle\cliffordconjugate{\Psi}\Psi\rangle
=\tfrac12(s_1^2+\cdots+s_8^2)
\end{multlined}
\end{tabular}
\caption{Real spinor table for  GAs with vector space dimension $n=3$}\label{spt3}
\end{table}
\vskip 10pt

\begin{table}[H]
\begin{tabular}{l|>{$}l<{$}>{$}l<{$}>{$}l<{$}}
& \cl{4}{0}[\bbH(2)]& \cl{3}{1}[\bbR(4)]
\\ \hline
1&
\begin{multlined}[t][0.1\columnwidth]
\{P_1=\tfrac12 (1 + \e{1}),
 P_2=\tfrac12(1 - \e{1})
 \}
\end{multlined}
&
\begin{multlined}[t][0.1\columnwidth]
\{P_1=\tfrac14 (1 + \e{1})(1 + \e{24}),
 P_2=(+,-), P_3=(-,+),
  P_4=(-,-)
\}
\end{multlined}
\\
2&
\begin{multlined}[t][0.1\columnwidth]
  \{P_1 1 P_1, P_1 \e{23} P_1, P_1 \e{24} P_1, P_1 \e{34} P_1 \}
\end{multlined}
&
\begin{multlined}[t][0.1\columnwidth]
  \{P_1 1 P_1\}
\end{multlined}
\\
3&
\begin{multlined}[t][0.1\columnwidth]
\{P_1, \e{2}P_1\}
\end{multlined}
&
\begin{multlined}[t][0.1\columnwidth]
\{P_1, \e{2}P_1,\e{3}P_1,\e{23}P_1\}
\end{multlined}
\\
4&
\begin{multlined}[t][0.1\columnwidth]
\{E_{11} - E_{22},
 E_{12} + E_{21},
-q_{1}E_{12} + q_{1}E_{21},\\[-3ex]
-q_{2}E_{12} + q_{2}E_{21}\}
\end{multlined}
&
\begin{multlined}[t][0.1\columnwidth]
\{E_{11} - E_{22} - E_{33} + E_{44},
E_{12} + E_{21} + E_{34} + E_{43},\\[-3ex]
E_{13} - E_{24} + E_{31} - E_{42},
 -E_{12} + E_{21} - E_{34} + E_{43}
\}
\end{multlined}
\\
5&
\begin{multlined}[t][0.1\columnwidth]
 \Psi=(s_{1}  + s_{2}\e{23} + s_{3}\e{24} + s_{4}\e{34} + s_{5}\e{2} - s_{6}\e{3}\\[-3ex] - s_{7}\e{4} + s_{8}\e{234})P_1
\end{multlined}
&
\begin{multlined}[t][0.1\columnwidth]
\Psi=(s_{1} + s_{2}\e{2} + s_{3}\e{3} + s_{4}\e{23}) P_1
\end{multlined}\\
6&
\begin{multlined}[t][0.1\columnwidth]
(s_{1} + s_{2}q_{1} + s_{3}q_{2} - s_{4}q_{3})E_{11}
+(s_{5} - s_{6}q_{1} - s_{7}q_{2}\\[-3ex] - s_{8}q_{3})E_{21}
\end{multlined}
&
s_{1}E_{11} +s_{2} E_{21} +s_{3} E_{31} +s_{4} E_{41}
\\
7&
\langle\widetilde{\Psi}\Psi\rangle =\tfrac12(s_1^2+\cdots+s_8^2)
&
\begin{multlined}[t][0.1\columnwidth]
-\langle\e{123}\widetilde{\Psi}\e{123}\Psi\rangle=
\tfrac14(s_1^2+\cdots+s_4^2)
\end{multlined}
\end{tabular}
\vskip 10pt

\begin{tabular}{l|>{$}l<{$}>{$}l<{$}>{$}l<{$}}
& \cl{2}{2}[\bbR(4)]&  \cl{1}{3}[\bbH(2)]
\\ \hline
1&
\begin{multlined}[t][0.1\columnwidth]
\{P_1=\tfrac14 (1 + \e{1})(1 + \e{23}),
 P_2=(+,-),
  P_3=(-,+),\\[-3ex]
  P_4=(-,-)
 \}
\end{multlined}
&
\begin{multlined}[t][0.1\columnwidth]
\{P_1=\tfrac12 (1 + \e{1}),
 P_2=\tfrac12 (1 - \e{1})
 \}
\end{multlined}
\\
2&
\begin{multlined}[t][0.1\columnwidth]
  \{P_1 1 P_1\}
\end{multlined}
&
\begin{multlined}[t][0.1\columnwidth]
\{P_1 1 P_1, P_1 \e{23} P_1, P_1 \e{24} P_1, P_1 \e{34} P_1 \}
\end{multlined}
\\
3&
\begin{multlined}[t][0.1\columnwidth]
\{P_1,\e{2}P_1,\e{4}P_1,\e{24}P_1\}
\end{multlined}
&
\begin{multlined}[t][0.1\columnwidth]
\{P_1, \e{2} P_1\}
\end{multlined}
\\
4&
\begin{multlined}[t][0.1\columnwidth]
\{E_{11} - E_{22} - E_{33} + E_{44},
E_{12} + E_{21} + E_{34} + E_{43},\\[-3ex]
-E_{12} + E_{21} - E_{34} + E_{43},
 -E_{13} + E_{24} + E_{31} - E_{42}
\}
\end{multlined}
&
\begin{multlined}[t][0.1\columnwidth]
\{E_{11} - E_{22},
  -E_{12} + E_{21},
-q_{1}E_{12} - q_{1}E_{21},
-q_{2}E_{12} - q_{2}E_{21}\}
\end{multlined}
\\
5&
\begin{multlined}[t][0.1\columnwidth]
\Psi=(s_{1} + s_{2}\e{2} + s_{3}\e{4} + s_{4}\e{24})P_1
\end{multlined}
&
\begin{multlined}[t][0.1\columnwidth]
\Psi=(s_{1} + s_{2}\e{23} + s_{3}\e{24} + s_{4}\e{34} + s_{5}\e{2} + s_{6}\e{3} + s_{7}\e{4} + s_{8}\e{234})P_1
\end{multlined}
\\
6&
s_{1}E_{11} +s_{2} E_{21} +s_{3} E_{31} +s_{4} E_{41}
&
\begin{multlined}[t][0.1\columnwidth]
(s_{1} + s_{2}q_{1}  + s_{3}q_{2} + s_{4}q_{3} )E_{11}
+ (s_{5} - s_{6}q_{1}  - s_{7}q_{2}  + s_{8}q_{3})E_{21}
\end{multlined}
\\
7&
\begin{multlined}[t][0.1\columnwidth]
-\langle\e{34}\widetilde{\Psi}\e{34}\Psi\rangle=\tfrac14(s_1^2+\cdots+s_4^2)
\end{multlined}
&
\begin{multlined}[t][0.1\columnwidth]
\langle\e{1}\widetilde{\Psi}\e{1}\Psi\rangle
=\tfrac12(s_1^2+\cdots+s_8^2)
\end{multlined}
\end{tabular}
\vskip 10pt

\begin{tabular}{l|>{$}l<{$}>{$}l<{$}>{$}l<{$}>{$}l<{$}}
& \cl{0}{4}[\bbH(2)]
\\ \hline
1&
\begin{multlined}[t][0.1\columnwidth]
\{P_1=\tfrac12 (1 + \e{123}),
 P_2=\tfrac12 (1 - \e{123})
 \}
\end{multlined}
&
\\
2&
\begin{multlined}[t][0.1\columnwidth]
\{P_1 1 P_1, P_1 \e{1} P_1, P_1 \e{2} P_1, P_1 \e{3} P_1 \}
\end{multlined}
&
\\
3&
\begin{multlined}[t][0.1\columnwidth]
  \{\e{3} P_1, \e{34} P_1\}
\end{multlined}
&
\\
4&
\begin{multlined}[t][0.1\columnwidth]
  \{-q_{1}E_{11} + q_{1}E_{22},
-q_{2}E_{11} + q_{2} E_{22},
-q_{3}E_{11}+q_{3}E_{22},
 E_{12} - E_{21}
\}
\end{multlined}
&
\\
5&
\begin{multlined}[t][0.1\columnwidth]
\Psi=(- s_{1} -s_{2}\e{1} + s_{3}\e{2} + s_{4}\e{3}  - s_{5}\e{4} + s_{6}\e{1,4} - s_{7}\e{24} + s_{8}\e{34} )P_1
\end{multlined}
&
\\
6&
\begin{multlined}[t][0.1\columnwidth]
(-s_{1} + s_{2}q_{1} - s_{3}q_{2} - s_{4} q_{3})E_{11}
+ (s_{5}- s_{6}q_{1} + s_{7}q_{2} - s_{8}q_{3})E_{21}
\end{multlined}
&
\\
7&
\begin{multlined}[t][0.1\columnwidth]
\langle\cliffordconjugate{\Psi}\Psi\rangle=\tfrac12(s_1^2+\cdots+s_8^2)
\end{multlined}
&
\end{tabular}
\caption{Real spinor tables for  GAs with vector space dimension $n=4$}\label{spt4}
\end{table}
\vskip 10pt

\begin{table}[H]
\begin{tabular}{l|>{$}l<{$}>{$}l<{$}>{$}l<{$}}
& \cl{5}{0}[{^2\bbH(2)}]& \cl{4}{1}[\bbC(4)]
 \index{algebra!\cl{5}{0}}\index{algebra!\cl{4}{1}}
\\ \hline
1&
\begin{multlined}[t][0.1\columnwidth]
\{P_1=\tfrac14 (1 + \e{1})(1 + \e{2345}),
 P_2=\tfrac14 (1 + \e{1})(1 - \e{2345}),\\[-3ex]
 P_3=\gradeinverse{P_1}=\tfrac14 (1 - \e{1})(1 + \e{2345}),
 P_4=\tfrac14 (1 - \e{1})(1 - \e{2345})\}
\end{multlined}
&
\begin{multlined}[t][0.1\columnwidth]
\{P_1=\tfrac14 (1 + \e{1})(1 + \e{25}),
 P_2=\tfrac14 (1 + \e{1})(1 - \e{25}),\\[-3ex]
 P_3=\tfrac14 (1 - \e{1})(1 + \e{25}),
 P_4=\tfrac14 (1 - \e{1})(1 - \e{25})\}
\end{multlined}
\\
2&
\begin{multlined}[t][0.1\columnwidth]
  \{P_1 1 P_1,P_1 \e{23} P_1,P_1 \e{24} P_1, P_1 \e{25} P_1\}
\end{multlined}
&
\begin{multlined}[t][0.1\columnwidth]
\{P_1 1 P_1,P_1 \e{34} P_1\}
\end{multlined}
\\
3&
\begin{multlined}[t][0.1\columnwidth]
  \{\e{25}P_1,\e{5}P_1\}\cup \{\gradeinverse{\e{25}P_1},\gradeinverse{\e{5}P_1}\}
\end{multlined}
&
\begin{multlined}[t][0.1\columnwidth]
\{P_1,\e{2}P_1,\e{3}P_1,\e{23}P_1\}
\end{multlined}
\\
4&
\begin{multlined}[t][0.1\columnwidth]
\{E_{11} - E_{22} - E_{33} + E_{44},
E_{12} + E_{21} - E_{34} - E_{43},\\[-3ex]
q_{1}(E_{12} - E_{21}  - E_{34}+ E_{43}),
q_{2}(E_{12}- E_{21} - E_{34} + E_{43}),\\[-1ex]
q_{3}(-E_{12} + E_{21} + E_{34}  - E_{43})
\}
\end{multlined}
&
\begin{multlined}[t][0.1\columnwidth]
\{E_{11} - E_{22} - E_{33} + E_{44},
E_{12} + E_{21} + E_{34} + E_{43},\\[-3ex]
E_{13} - E_{24} + E_{31} - E_{42},
\ii(-E_{13} + E_{24} + E_{31} - E_{42}),\\[-1ex]
-E_{12} + E_{21} - E_{34} + E_{43}
\}
\end{multlined}
\\
5&
\begin{multlined}[t][0.1\columnwidth]
\Psi=  (-s_{1} + s_{2}\e{23} - s_{3}\e{24} + s_{4}\e{25}  + s_{5}\e{2}  - s_{6}\e{3} + s_{7}\e{4} + s_{8}\e{5} )P_1\\[-3ex]
+(-s_{9} + s_{10}\e{23} - s_{11}\e{24} + s_{12}\e{25} - s_{13}\e{2}\\[-1ex]+  s_{14}\e{3} - s_{15}\e{4} - s_{16}\e{5} )\gradeinverse{P_1}
\end{multlined}
&
\begin{multlined}[t][0.1\columnwidth]
\Psi= (s_{1} + s_{2}\e{34} + s_{3}\e{2} + s_{4}\e{234} \\[-3ex]
 \quad + s_{5}\e{3} - s_{6}\e{4} + s_{7}\e{23} - s_{8}\e{24})P_1
\end{multlined}
\\
6&
\begin{multlined}[t][0.1\columnwidth]
(-s_{1} -s_{2}q_{1}  + s_{3}q_{2} + s_{4}q_{3})E_{11}
+ (s_{5}+ s_{6}q_{1} - s_{7}q_{2} + s_{8}q_{3})E_{21}\\[-3ex]
+ (- s_{9} - s_{10}q_{1} + s_{11}q_{2} + s_{12}q_{3})E_{33}\\[-1ex]
+ ( s_{13}+s_{14}q_{1}  - s_{15}q_{2} + s_{16}q_{3})E_{43}
\end{multlined}
&
\begin{multlined}[t][0.1\columnwidth]
(s_{1} + \ii s_{2})E_{11}
+ (s_{3} + \ii s_{4})E_{21}
+ (s_{5} - \ii s_{6})E_{31}\\[-3ex]
+ (s_{7} - \ii s_{8})E_{41}
\end{multlined}
\\
7&
\begin{multlined}[t][0.1\columnwidth]
\langle\widetilde{\Psi}\Psi\rangle=\tfrac14(s_1^2+\cdots+s_{16}^2)
\end{multlined}
&
\begin{multlined}[t][0.1\columnwidth]
-\langle\e{5}\cliffordconjugate{\Psi}\e{5}\Psi\rangle=\tfrac14(s_1^2+\cdots+s_8^2)
\end{multlined}

\end{tabular}
\vskip 10pt

\begin{tabular}{l|>{$}l<{$}}
& \cl{3}{2}[{^2\bbR(4)}]
\\ \hline
1&
\begin{multlined}[t][0.1\columnwidth]
\{P_1=\tfrac18 (1 + \e{1})(1 + \e{24})(1 + \e{35}),
P_2=(+,+,-), P_3=(+,-,+), P_4=(+,-,-),
P_5=\gradeinverse{P_1}=(-,+,+),
P_6=(-,+,-),\\[-3ex]
P_7=(-,-,+), P_8=(-,-,-)
\}
\end{multlined}
\\
2&
\begin{multlined}[t][0.1\columnwidth]
\{P_1 1 P_1\}
\end{multlined}
\\
3&
\begin{multlined}[t][0.1\columnwidth]
  \{P_1, \e{2} P_1,\e{3} P_1,\e{23} P_1\}\cup \{\gradeinverse{P_1}, \gradeinverse{\e{2} P_1},\gradeinverse{\e{3} P_1}, \gradeinverse{\e{23} P_1}
\}
\end{multlined}
\\
4&
\begin{multlined}[t][0.1\columnwidth]
\{E_{11} - E_{22} - E_{33} + E_{44} - E_{55} + E_{66} + E_{77} - E_{88},
E_{12} + E_{21} + E_{34} + E_{43} - E_{56} - E_{65} - E_{78} - E_{87},\\[-3ex]
E_{13} - E_{24} + E_{31} - E_{42} - E_{57} + E_{68} - E_{75} + E_{86},
-E_{12} + E_{21} - E_{34} + E_{43} + E_{56} - E_{65} + E_{78} - E_{87},\\[-1ex]
-E_{13} + E_{24} + E_{31} - E_{42} + E_{57} - E_{68} - E_{75} + E_{86}
\}
\end{multlined}
\\
5&
\begin{multlined}[t][0.1\columnwidth]
\Psi= (s_{1} + s_{2}\e{2} + s_{3}\e{3} + s_{4}\e{23})P_1
+(s_{5} - s_{6}\e{2} - s_{7}\e{3} + s_{8}\e{23})\gradeinverse{P_1}
\end{multlined}
\\
6&
\begin{multlined}[t][0.1\columnwidth]
s_{1}E_{11} +s_{2} E_{21}+ s_{3} E_{31}+ s_{4} E_{41} +s_{5}E_{55}
+s_{6} E_{65}+ s_{7} E_{75}+ s_{8} E_{85}
\end{multlined}
\\
7&
\begin{multlined}[t][0.1\columnwidth]
-\langle\e{45}\reverse{\Psi}\e{45}\Psi\rangle=\tfrac18(s_1^2+\cdots+s_8^2)
\end{multlined}
\end{tabular}
\vskip 10pt

\begin{tabular}{l|>{$}l<{$}}
& \cl{2}{3}[\bbC(4)]
\\ \hline
1&
\begin{multlined}[t][0.1\columnwidth]
\{P_1=\tfrac14 (1 + \e{1})(1 + \e{23}),
P_2=(+,-), P_3=(-,+), P_4=(-,-)\}
\end{multlined}
\\
2&
\begin{multlined}[t][0.1\columnwidth]
\{P_1 1 P_1,P_1 \e{45} P_1\}
\end{multlined}
\\
3&
\begin{multlined}[t][0.1\columnwidth]
\{P_1, \e{2} P_1,\e{4} P_1,\e{24} P_1\}
\end{multlined}
\\
4&
\begin{multlined}[t][0.1\columnwidth]
\{E_{11} - E_{22} - E_{33} + E_{44},
E_{12} + E_{21} + E_{34} + E_{43},
-E_{12} + E_{21} - E_{34} + E_{43},
  -E_{13} + E_{24} + E_{31} - E_{42},\\[-3ex]
\ii(-E_{13} + E_{24} - E_{31} + E_{42})
\}
\end{multlined}
\\
5&
\begin{multlined}[t][0.1\columnwidth]
\Psi=(s_{1} + s_{2}\e{45} + s_{3}\e{2}+ s_{4}\e{245} + s_{5}\e{4}
+ s_{6}\e{5} + s_{7}\e{24} + s_{8}\e{25}) P_1
\end{multlined}
\\
6&
(s_{1} + \ii s_{2})E_{11} +(s_{3} + \ii s_{4}) E_{21} +(s_{5} - \ii s_{6}) E_{31} +(s_{7} - \ii s_{8}) E_{41}
\\
7&
\begin{multlined}[t][0.1\columnwidth]
-\langle\e{12}\cliffordconjugate{\Psi}\e{12}\Psi\rangle=\tfrac18(s_1^2+\cdots+s_8^2)
\end{multlined}
\end{tabular}
\vskip 10pt

\begin{tabular}{l|>{$}l<{$}}
& \cl{1}{4}[{^2\bbH(2)}]
\\ \hline
1&
\begin{multlined}[t][0.1\columnwidth]
\{P_1=\tfrac14 (1 + \e{1})(1 + \e{2345}),
P_2=(+,-), P_3=\gradeinverse{P_1}=(-,+), P_4=(-,-)\}
\end{multlined}
\\
2&
\begin{multlined}[t][0.1\columnwidth]
\{P_1 1 P_1, P_1 \e{23} P_1, P_1\e{24} P_1, P_1 \e{25} P_1\}
\end{multlined}
\\
3&
\begin{multlined}[t][0.1\columnwidth]
\{\e{25}P_1,\e{5}P_1\}\cup \{\gradeinverse{\e{25}P_1},\gradeinverse{\e{5}P_1}\}
\end{multlined}
\\
4&
\begin{multlined}[t][0.1\columnwidth]
\{E_{11} - E_{22} - E_{33} + E_{44},
E_{12} - E_{21} - E_{34} + E_{43},
q_{1}(-E_{12} - E_{21} + E_{34} + E_{43}),
q_{2}(-E_{12} - E_{21} + E_{34} + E_{43}),\\[-3ex]
q_{3}(-E_{12} - E_{21} + E_{34} + E_{43})
\}
\end{multlined}
\\
5&
\begin{multlined}[t][0.1\columnwidth]
\Psi=(-s_{1} - s_{2}\e{23} + s_{3}\e{24} + s_{4}\e{25} - s_{5}\e{2} + s_{6}\e{3} - s_{7}\e{4} + s_{8}\e{5})P_1\\[-3ex]
\quad+(-s_{9} -s_{10}\e{23} + s_{11}\e{24} + s_{12}\e{25} + s_{13}\e{2} - s_{14}\e{3} + s_{15}\e{4} - s_{16}\e{5})\gradeinverse{P_1}
\end{multlined}
\\
6&
\begin{multlined}[t][0.1\columnwidth]
(-s_{1} + s_{2}q_{1} - s_{3}q_{2} - s_{4}q_{3})E_{11} + (s_{5}- s_{6}q_{1} + s_{7}q_{2}  - s_{8}q_{3}  ) E_{21}
+ (-s_{9} + s_{10}q_{1}  - s_{11}q_{2} - s_{12}q_{3})E_{33}\\[-3ex] + (s_{13}- s_{14}q_{1} + s_{15}q_{2} - s_{16}q_{3})E_{43}
\end{multlined}
\\
7&
\begin{multlined}[t][0.1\columnwidth]
\langle\e{1}\reverse{\Psi}\e{1}\Psi\rangle=
\tfrac18(s_1^2+\cdots+s_{16}^2)
\end{multlined}
\end{tabular}
\vskip 10pt

\begin{tabular}{l|>{$}l<{$}}
& \cl{0}{5}[\bbC(4)]
\\ \hline
1&
\begin{multlined}[t][0.1\columnwidth]
\{P_1=\tfrac14 (1 + \e{123})(1 + \e{145}),
P_2=(+,-), P_3=(-,+), P_4=(-,-)\}
\end{multlined}
\\
2&
\begin{multlined}[t][0.1\columnwidth]
  \{P_1 1 P_1, P_1 \e{1} P_1\}
\end{multlined}
\\
3&
\begin{multlined}[t][0.1\columnwidth]
\{P_1,\e{2} P_1, \e{4} P_1, \e{24} P_1\}
\end{multlined}
\\
4&
\begin{multlined}[t][0.1\columnwidth]
\{\ii(E_{11} - E_{22} - E_{33} + E_{44}),
-E_{12} + E_{21} - E_{34} + E_{43},
\ii(E_{12} + E_{21} + E_{34} + E_{43}),
-E_{13} + E_{24} + E_{31} - E_{42},\\[-3ex]
\ii(E_{13} - E_{24} + E_{31} - E_{42})
\}
\end{multlined}
\\
5&
\begin{multlined}[t][0.1\columnwidth]
 \Psi= (s_{1} + s_{2}\e{1} + s_{3}\e{2}- s_{4}\e{3} + s_{5}\e{4} - s_{6}\e{5} + s_{7}\e{24} + s_{8}\e{25})P_1
\end{multlined}
\\
6&
(s_{1} + \ii s_{2})E_{11} + (s_{3} - \ii s_{4})E_{21} + (s_{5} - \ii s_{6})E_{31} + (s_{7} + \ii s_{8})E_{41}
\\
7&
\begin{multlined}[t][0.1\columnwidth]
\langle\cliffordconjugate{\Psi}\Psi\rangle=\tfrac14(s_1^2+\cdots+s_{8}^2)
\end{multlined}
\end{tabular}
\caption{Real spinor tables for  GAs with vector space dimension $n=5$}\label{spt5}
\end{table}\vskip 10pt

\begin{table}[H]
\begin{tabular}{l|>{$}l<{$}>{$}l<{$}}
& \cl{6}{0}[{\bbH(4)}]\index{algebra!\cl{6}{0}}
\\ \hline
1&
\begin{multlined}[t][0.1\columnwidth]
\{P_1=\tfrac14 (1 + \e{1})(1 + \e{2345}),
P_2=(+,-), P_3=\gradeinverse{P_1}=(-,+), P_4=(-,-)\}
\end{multlined}
\\
2&
\begin{multlined}[t][0.1\columnwidth]
  \{P_1 1 P_1, P_1 \e{23} P_1, P_1 \e{24} P_1, P_1 \e{25} P_1\}
\end{multlined}
\\
3&
\begin{multlined}[t][0.1\columnwidth]
  \{\e{25} P_1,\e{5} P_1,\e{256} P_1,\e{56} P_1\}
\end{multlined}
\\
4&
\begin{multlined}[t][0.1\columnwidth]
\{E_{11} - E_{22} - E_{33} + E_{44},
E_{12} + E_{21} + E_{34} + E_{43},
q_{1}(E_{12} - E_{21}+ E_{34} - E_{43}),
q_{2}(E_{12} - E_{21}  + E_{34} - E_{43}),\\[-3ex]
q_{3}(-E_{12}+ E_{21} - E_{34}+ E_{43}),
E_{13} - E_{24} + E_{31} - E_{42}
\}
\end{multlined}
\\
5&
\begin{multlined}[t][0.1\columnwidth]
\Psi=(-s_{1} +s_{2}\e{23} - s_{3}\e{24} + s_{4}\e{25} + s_{5}\e{2} - s_{6}\e{3} + s_{7}\e{4} + s_{8}\e{5}\\[-3ex]\quad - s_{9}\e{6} + s_{10}\e{236} - s_{11}\e{246} + s_{12}\e{256} + s_{13}\e{26} - s_{14}\e{36} + s_{15}\e{46} + s_{16}\e{56} )P_1
\end{multlined}
\\
6&
\begin{multlined}[t][0.1\columnwidth]
(-s_{1} -s_{2}q_{1}  + s_{3}q_{2}  + s_{4}q_{3} )E_{11} + (s_{5}+ s_{6}q_{1} - s_{7}q_{2} + s_{8}q_{3} )E_{21} + (-s_{9} -s_{10}q_{1} + s_{11}q_{2} + s_{12}q_{3} )E_{31}\\[-3ex] + ( s_{13}+s_{14}q_{1}  - s_{15}q_{2} + s_{16}q_{3})E_{41}
\end{multlined}
\\
7&
\begin{multlined}[t][0.1\columnwidth]
\langle\reverse{\Psi}\Psi\rangle=\tfrac14(s_1^2+\cdots+s_{16}^2)
\end{multlined}
\end{tabular}
\vskip 10pt

\begin{tabular}{l|>{$}l<{$}}
 & \cl{5}{1}[\bbH(4)]
\\ \hline
1&
\begin{multlined}[t][0.1\columnwidth]
\{P_1=\tfrac14 (1 + \e{1})(1 + \e{26}),
P_2=(+,-), P_3=(-,+), P_4=(-,-)\}
\end{multlined}
\\
2&
\begin{multlined}[t][0.1\columnwidth]
  \{P_1 1 P_1, P_1 \e{34} P_1, P_1 \e{35} P_1, P_1 \e{45} P_1\}
\end{multlined}
\\
3&
\begin{multlined}[t][0.1\columnwidth]
\{P_1,\e{2}P_1,\e{3}P_1,\e{23}P_1\}
\end{multlined}
\\
4&
\begin{multlined}[t][0.1\columnwidth]
\{E_{11} - E_{22} - E_{33} + E_{44},
 E_{12} + E_{21} + E_{34} + E_{43},
E_{13} - E_{24} + E_{31} - E_{42},
q_{1}(-E_{13} + E_{24} + E_{31}- E_{42}),\\[-3ex]
q_{2}(-E_{13} + E_{24} + E_{31}- E_{42}),
-E_{12} + E_{21} - E_{34} + E_{43}
\}
\end{multlined}
\\
5&
\begin{multlined}[t][0.1\columnwidth]
\Psi=(s_{1}+ s_{2}\e{34} + s_{3}\e{35} + s_{4}\e{45} + s_{5}\e{2} + s_{6}\e{234} + s_{7}\e{235} + s_{8}\e{245} \\[-3ex]\quad+ s_{9}\e{3} - s_{10}\e{4} - s_{11}\e{5} + s_{12}\e{345}  + s_{13}\e{23} - s_{14}\e{24} - s_{15}\e{25} + s_{16}\e{2345}) P_1
\end{multlined}
\\
6&
\begin{multlined}[t][0.1\columnwidth]
(s_{1} + s_{2}q_{1} + s_{3}q_{2} - s_{4}q_{3})E_{11} + (s_{5} + s_{6}q_{1} + s_{7}q_{2} - s_{8}q_{3} )E_{21} + (s_{9} - s_{10}q_{1} - s_{11}q_{2} - s_{12}q_{3} )E_{31}\\[-3ex] + (s_{13} - s_{14}q_{1}  - s_{15}q_{2}  - s_{16}q_{3})E_{41}
\end{multlined}
\\
7&
\begin{multlined}[t][0.1\columnwidth]
-\langle\e{6}\cliffordconjugate{\Psi}\e{6}\Psi\rangle=\tfrac14(s_1^2+\cdots+s_{16}^2)
\end{multlined}
\end{tabular}
\vskip 10pt

\begin{tabular}{l|>{$}l<{$}}
 &  \cl{4}{2}[{\bbR(8)}]
\\ \hline
1&
\begin{multlined}[t][0.1\columnwidth]
\{P_1=\tfrac18 (1 + \e{1})(1 + \e{25})(1 + \e{36}),
P_2=(+,+,-), P_3=(+,-,+), P_4=(+,-,-),
P_5=(-,+,+),
P_6=(-,+,-),\\[-3ex] P_7=(-,-,+), P_8=(-,-,-)
\}
\end{multlined}
\\
2&
\begin{multlined}[t][0.1\columnwidth]
  \{P_1 1 P_1\}
\end{multlined}
\\
3&
\begin{multlined}[t][0.1\columnwidth]
  \{P_1,\e{2}P_1,\e{3}P_1,\e{23}P_1,\e{4}P_1,\e{24}P_1,\e{34}P_1,\e{234}P_1\}
\end{multlined}
\\
4&
\begin{multlined}[t][0.1\columnwidth]
\{E_{11} - E_{22} - E_{33} + E_{44} - E_{55} + E_{66} + E_{77} - E_{88},
E_{12} + E_{21} + E_{34} + E_{43} + E_{56} + E_{65} + E_{78} + E_{87},\\[-3ex]
E_{13} - E_{24} + E_{31} - E_{42} + E_{57} - E_{68} + E_{75} - E_{86},
E_{15} - E_{26} - E_{37} + E_{48} + E_{51} - E_{62} - E_{73} + E_{84},\\[-1ex]
-E_{12} + E_{21} - E_{34} + E_{43} - E_{56} + E_{65} - E_{78} + E_{87},
-E_{13} + E_{24} + E_{31} - E_{42} - E_{57} + E_{68} + E_{75} - E_{86}
\}
\end{multlined}
\\
5&
\begin{multlined}[t][0.1\columnwidth]
\Psi=(s_{1} + s_{2}\e{2} + s_{3}\e{3}  + s_{4}\e{23}+ s_{5}\e{4} +
s_{6}\e{24} + s_{7}\e{34} + s_{8}\e{234})P_1
\end{multlined}
\\
6&
\begin{multlined}[t][0.1\columnwidth]
s_{1}E_{11} +s_{2} E_{21}+ s_{3} E_{31}+ s_{4} E_{41}
+s_{5}E_{51} +s_{6} E_{61}+ s_{7} E_{71}+ s_{8} E_{81}
\end{multlined}
\\
7&
\begin{multlined}[t][0.1\columnwidth]
 -\langle\e{56}\reverse{\Psi}\e{56}\Psi\rangle=\tfrac18(s_1^2+\cdots+s_{8}^2)
\end{multlined}
\end{tabular}
\vskip 10pt

\begin{tabular}{l|>{$}l<{$}}
 & \cl{3}{3}[\bbR(8)]
\\ \hline
1&
\begin{multlined}[t][0.1\columnwidth]
\{P_1=\tfrac18 (1 + \e{1})(1 + \e{24})(1 + \e{35}),
P_2=(+,+,-), P_3=(+,-,+), P_4=(+,-,-),
P_5=(-,+,+),
P_6=(-,+,-),\\[-3ex] P_7=(-,-,+), P_8=(-,-,-)
\}
\end{multlined}
\\
2&
\begin{multlined}[t][0.1\columnwidth]
\{P_1 1 P_1 \}
\end{multlined}
\\
3&
\begin{multlined}[t][0.1\columnwidth]
\{P_1,\e{2}P_1,\e{3}P_1,\e{23}P_1,\e{6}P_1,\e{26}P_1,\e{36}P_1,\e{236}P_1\}
\end{multlined}
\\
4&
\begin{multlined}[t][0.1\columnwidth]
\{E_{11} - E_{22} - E_{33} + E_{44} - E_{55} + E_{66} + E_{77} - E_{88},
E_{12} + E_{21} + E_{34} + E_{43} + E_{56} + E_{65} + E_{78} + E_{87},\\[-3ex]
E_{13} - E_{24} + E_{31} - E_{42} + E_{57} - E_{68} + E_{75} - E_{86},
-E_{12} + E_{21} - E_{34} + E_{43} - E_{56} + E_{65} - E_{78} + E_{87},\\[-1ex]
-E_{13} + E_{24} + E_{31} - E_{42} - E_{57} + E_{68} + E_{75} - E_{86},
-E_{15} + E_{26} + E_{37} - E_{48} + E_{51} - E_{62} - E_{73} + E_{84}
\}
\end{multlined}
\\
5&
\begin{multlined}[t][0.1\columnwidth]
\Psi=(s_{1} + s_{2}\e{2} + s_{3}\e{3}+ s_{4}\e{23} + s_{5}\e{6}  +
s_{6}\e{26} + s_{7}\e{36} + s_{8}\e{236})P_1
\end{multlined}
\\
6&
\begin{multlined}[t][0.1\columnwidth]
s_{1}E_{11} +s_{2} E_{21}+ s_{3} E_{31}+ s_{4} E_{41} +s_{5}E_{51}
+s_{6} E_{61}+ s_{7} E_{71}+ s_{8} E_{81}
\end{multlined}
\\
7&
\begin{multlined}[t][0.1\columnwidth]
 -\langle\e{123}\reverse{\Psi}\e{123}\Psi\rangle=\tfrac18(s_1^2+\cdots+s_{8}^2)
\end{multlined}
\end{tabular}
\vskip 10pt

\begin{tabular}{l|>{$}l<{$}}
& \cl{2}{4}[\bbH(4)]
\\ \hline
1&
\begin{multlined}[t][0.1\columnwidth]
\{P_1=\tfrac14 (1 + \e{1})(1 + \e{23}),
P_2=(+,-), P_3=(-,+), P_4=(-,-)\}
\end{multlined}
\\
2&
\begin{multlined}[t][0.1\columnwidth]
 \{P_1 1 P_1, P_1 \e{45} P_1, P_1 \e{46} P_1, P_1 \e{56} P_1\}
\end{multlined}
\\
3&
\begin{multlined}[t][0.1\columnwidth]
\{P_1,\e{2}P_1,\e{4}P_1,\e{24}P_1\}
\end{multlined}
\\
4&
\begin{multlined}[t][0.1\columnwidth]
\{E_{11} - E_{22} - E_{33} + E_{44},
E_{12} + E_{21} + E_{34} + E_{43},
-E_{12} + E_{21} - E_{34} + E_{43},
-E_{13} + E_{24} + E_{31} - E_{42},\\[-3ex]
q_{1}(-E_{13}+ E_{24}- E_{31}+ E_{42}),
q_{2}(-E_{13} + E_{24}- E_{31} + E_{42})
\}
\end{multlined}
\\
5&
\begin{multlined}[t][0.1\columnwidth]
\Psi=(s_{1}+ s_{2}\e{45} + s_{3}\e{46} + s_{4}\e{56} + s_{5}\e{2}  + s_{6}\e{245} + s_{7}\e{246} + s_{8}\e{256} + s_{9}\e{4} + s_{10}\e{5} + s_{11}\e{6}  + s_{12}\e{456}+ s_{13}\e{24} + s_{14}\e{25}\\[-3ex] + s_{15}\e{26}  + s_{16}\e{2456})P_1
\end{multlined}
\\
6&
\begin{multlined}[t][0.1\columnwidth]
(s_{1} + s_{2}q_{1}  + s_{3}q_{2}  + s_{4}q_{3} )E_{11} + (s_{5} + s_{6}q_{1}  + s_{7}q_{2}  + s_{8}q_{3} )E_{21} \\[-3ex]+ (s_{9} - s_{10}q_{1} - s_{11}q_{2}  + s_{12}q_{3} )E_{31} + (s_{13} - s_{14}q_{1}  - s_{15}q_{2}  + s_{16}q_{3} )E_{41}
\end{multlined}
\\
7&
\begin{multlined}[t][0.1\columnwidth]
-\langle\e{12}\cliffordconjugate{\Psi}\e{12}\Psi\rangle=\tfrac14(s_1^2+\cdots+s_{16}^2)
\end{multlined}
\end{tabular}
\caption{Real spinor tables for GAs $\cl{6}{0}$, $\cl{5}{1}$, $\cl{4}{2}$, $\cl{3}{3}$, and  $\cl{2}{4}$; $n=6$}\label{spt6a}
\end{table}
\vskip 10pt

\begin{table}[H]
\begin{tabular}{l|>{$}l<{$}}
& \cl{1}{5}[\bbH(4)]
\\ \hline
1&
\begin{multlined}[t][0.1\columnwidth]
\{P_1=\tfrac14 (1 + \e{1})(1 + \e{2345}),
P_2=(+,-), P_3=(-,+), P_4=(-,-)\}
\end{multlined}
\\
2&
\begin{multlined}[t][0.1\columnwidth]
 \{P_1 1 P_1, P_1 \e{23} P_1, P_1 \e{24} P_1, P_1 \e{25} P_1\}
\end{multlined}
\\
3&
\begin{multlined}[t][0.1\columnwidth]
\{\e{25}P_1,\e{5}P_1,\e{256}P_1,\e{56}P_1\}
\end{multlined}
\\
4&
\begin{multlined}[t][0.1\columnwidth]
\{E_{11} - E_{22} - E_{33} + E_{44},
E_{12} - E_{21} + E_{34} - E_{43},
q_{1}(-E_{12} - E_{21} - E_{34}- E_{43}),
q_{2}(-E_{12} - E_{21} - E_{34} - E_{43}),\\[-3ex]
q_{3}(-E_{12} - E_{21} - E_{34}- E_{43}),
-E_{13} + E_{24} + E_{31} - E_{42}
\}
\end{multlined}
\\
5&
\begin{multlined}[t][0.1\columnwidth]
\Psi=(- s_{1} -s_{2}\e{23} + s_{3}\e{24} + s_{4}\e{25} - s_{5}\e{2} + s_{6}\e{3} - s_{7}\e{4} + s_{8}\e{5} - s_{9}\e{6} - s_{10}\e{236} + s_{11}\e{246} + s_{12}\e{256} - s_{13}\e{26} + s_{14}\e{36}\\[-3ex]- s_{15}\e{46} + s_{16}\e{56})P_1
\end{multlined}
\\
6&
\begin{multlined}[t][0.1\columnwidth]
  (-s_{1} + s_{2}q_{1} - s_{3}q_{2} - s_{4}q_{3})E_{11} + (s_{5}- s_{6}q_{1} + s_{7}q_{2}- s_{8}q_{3} )E_{21} \\[-3ex]+ (-s_{9} +s_{10}q_{1} - s_{11}q_{2} - s_{12}q_{3} )E_{31} +(s_{13} - s_{14}q_{1} + s_{15}q_{2}  - s_{16}q_{3}) E_{41}
\end{multlined}
\\
7&
\begin{multlined}[t][0.1\columnwidth]
\langle\e{1}\reverse{\Psi}\e{1}\Psi\rangle=\tfrac14(s_1^2+\cdots+s_{16}^2)
\end{multlined}
\end{tabular}
\vskip 10pt

\begin{tabular}{l|>{$}l<{$}}
& \cl{0}{6}[\bbR(8)]\index{algebra!\cl{0}{6}}
\\ \hline
1&
\begin{multlined}[t][0.1\columnwidth]
\{P_1=\tfrac18 (1 + \e{123})(1 + \e{145})(1 + \e{246}),
P_2=(+,+,-), P_3=(+,-,+), P_4=(+,-,-),
P_5=(-,+,+),
P_6=(-,+,-),\\[-3ex] P_7=(-,-,+), P_8=(-,-,-)
\}
\end{multlined}
\\
2&
\begin{multlined}[t][0.1\columnwidth]
  \{P_1 1 P_1 \}
\end{multlined}
\\
3&
\begin{multlined}[t][0.1\columnwidth]
\{P_1,\e{1}P_1,\e{2}P_1,\e{3}P_1,\e{4}P_1,\e{5}P_1,\e{6}P_1,\e{16}P_1\}
\end{multlined}
\\
4&
\begin{multlined}[t][0.1\columnwidth]
\{-E_{12} + E_{21} + E_{34} - E_{43} + E_{56} - E_{65} - E_{78} + E_{87},
-E_{13} - E_{24} + E_{31} + E_{42} + E_{57} + E_{68} - E_{75} - E_{86},\\[-3ex]
 -E_{14} + E_{23} - E_{32} + E_{41} + E_{58} - E_{67} + E_{76} - E_{85},
-E_{15} - E_{26} - E_{37} - E_{48} + E_{51} + E_{62} + E_{73} + E_{84},\\[-1ex]
-E_{16} + E_{25} - E_{38} + E_{47} - E_{52} + E_{61} - E_{74} + E_{83},
-E_{17} + E_{28} + E_{35} - E_{46} - E_{53} + E_{64} + E_{71} - E_{82}
\}
\end{multlined}
\\
5&
\begin{multlined}[t][0.1\columnwidth]
\Psi=(s_{1} + s_{2}\e{1} + s_{3}\e{2}  + s_{4}\e{3} + s_{5}\e{4} +
s_{6}\e{5} + s_{7}\e{6} + s_{8}\e{16})P_1
\end{multlined}
\\
6&
\begin{multlined}[t][0.1\columnwidth]
s_{1}E_{11} +s_{2} E_{21}+ s_{3} E_{31}+ s_{4} E_{41}
+s_{5}E_{51} +s_{6} E_{61}+ s_{7} E_{71}+ s_{8} E_{81}
\end{multlined}
\\
7&
\langle\cliffordconjugate{\Psi}\Psi\rangle=
\tfrac18(s_1^2+\cdots+s_{8}^2)
\end{tabular}
\caption{Real spinor tables for GAs $\cl{1}{5}$ and $\cl{0}{6}$; $n=6$}\label{spt6b}
\end{table}
\vskip 10pt

\end{flushleft}

\subsection{Example of construction of the real spinor data table for \cl{2}{2} algebra}.
\label{CL22exampleB}
\begin{itemize}

  \item[1.]  \textit{Primitive idempotents $P$}.
    In \cl{2}{2} we need to choose a single pair of mutually commuting blades $(\e{T_i},\e{T_j})$,
    \[
      \begin{array}{l}
	\{
      (\e{1}, \e{23}), (\e{1},\e{24}), ( \e{1},\e{123}), ( \e{1} , \e{124} ), ( \e{2} , \e{13} ),
    (\e{2} , \e{14}), ( \e{2} , \e{123} ), (\e{2}, \e{124}), 
	(\e{13} , \e{24} ), ( \e{13} , \e{123} ), \\ 
	\  ( \e{13} , \e{1234} ), (\e{14}, \e{23} ), ( \e{14} , \e{124} ),
    (\e{14} , \e{1234} ), ( \e{23} , \e{123} ), ( \e{23} , \e{1234}), ( \e{24} , \e{124}), ( \e{24} , \e{1234} )
  \}
      \end{array}
    \] ordered in InvDeg[Lex] that all squares to $+1$ and can be used in starting the construction of a spinor (see Table~\ref{spt4}). Each pair is mutually annihilating: $P_iP_j=P_jP_i=0$ if $i\ne j$, and $i,j=1,2$. From all possible sets we always will use the first one in the InvDeg[Lex] list, which means that we usually choose $\e{1}$ direction as a quantization axis. Exceptions are anti-euclidean algebras $\cl{0}{q}$. In accord with Table~\ref{spt4} we  choose the pair, $(\e{1},\e{23})$ and the primitive idempotent $P_1\to
P=\frac{1}{4}(1+\e{1})(1+\e{23})=\frac{1}{4}(1+\e{1}+\e{23}+\e{123})\equiv(+,+)$.\newline 
    \hphantom{\quad} Once the idempotent $P$ is chosen, multiply all basis elements of \cl{2}{2} by $P$ from right,
$S=\{\cl{2}{2}\}P$. Elements that differ by sign will
be treated as equivalent, therefore, only a single copy (positive)
is used in the following. One gets that the left ideal $S$
consists of four different members,
$S=\{P,\e{2}P,\e{4}P,\e{24}P\}=\{\tfrac{1}{4}(1+\e{1}+\e{23}+\e{123}),\tfrac{1}{4}(\e{2}+\e{3}-\e{12}-\e{13}),
\tfrac{1}{4} (\e{4}-\e{14}+\e{234}-\e{1234}),
\tfrac{1}{4}(\e{24}+\e{34}+\e{124}+\e{134})\}$.

\item[2.] \textit{Two-sided ideal (division ring)}. Find a list of
two-sided ideal (division ring) $K$ by multiplying the obtained
left ideal elements by same idempotent from  left,
$K=P\, S=P\{\cl{2}{2}\}P$. The division ring of $\cl{2}{2}$ contains single 
element 
    $K=\{K(1)\}=P1P=\tfrac{1}{4}(1+\e{1}+\e{23}+\e{123})$. Being an
idempotent, $K(1)$ represents a unit element of the real field,
$K(1)^2=K(1)$, which echoes the property $1*1=1$. Similarly, if
    $K$ list would contain two elements (remind that elements that differ by sign are considered as same elements), then the other element will play the role of an imaginary unit. And, if
$K$  would contain four different elements it will be isomorphic
to quaternion ring. The list of elements of the
two sided ideal should be sorted  according to their interpretation.
In particular, the element that plays the role of real unit
always appears first in the list. It is followed by 
elements that play the role of imaginary units. If $K$ has four elements,  
 the last three positions are occupied by non-commutative elements and (after proper reordering an sign change, when needed) can be interpreted as quaternion units $q_1, q_2$ and $q_3$. 

\item[3.] \textit{Ideal basis}.  The left ideal basis is obtained by
comparing all the elements of ideal (starting from the first) and
consequently dropping out all elements that can be obtained from
previous elements by multiplying them by any division ring
element. Since element choice may depend on its position in the
ideal list, the element ordering plays an important role. It is convenient to sort the elements of ideal basis by `RevLex` order what ensures that in spinor matrix
representation all columns will be zero except the first one. 
Other orderings in ideal basis will give different matrix
representations of the spinor.  In $\cl{2}{2}$ case the division ring contains a single element,
therefore, the ideal basis coincides with the ideal itself. 

\item[4.] \textit{Matrix rep of GA basis vectors}. Once the four-component ideal basis $S=\{S(1),S(2),S(3),S(4)\}$ is determined one can compute a matrix rep of basis vectors in algebraic spinor basis. The matrix  entry $(ij)$ can be calculated by $E_{ij}(\e{k})=\reverse{S}(i)^{\sharp}\e{k}S(j)$, where
$S(i)^{\sharp}$ is the ideal element in reciprocal basis. For example, if $S(1)=\tfrac{1}{4} (1+\e{1}+\e{23}+\e{123})$ then
$S(1)^\sharp=\tfrac{1}{4} (1+\e{}^{1}+\e{}^{23}+\e{}^{123})=\tfrac{1}{4} (1+\e{1}-\e{23}-\e{123})$, because in \cl{2}{2} the orthonormal of
reciprocal generators (basis) are $\e{}^1=\e{1}$, $\e{}^2=\e{2}$, $\e{}^3=-\e{3}$, and $\e{}^4=-\e{4}$. For each pair of indices
$(ij)$ in the ideal basis we obtain matrix entries $E_{ij}(\e{k})$ at fixed basis vector $\e{k}$. In our case $i,j=1,2,3,4$ and $k=1,2,3,4$. The
basis matrices $\hat{e}_1$, $\hat{e}_2$, $\hat{e}_3$, and $\hat{e}_4$ calculated with $E_{ij}(\e{k})$ then are, respectively,
\[\kern-2em
\hat{e}_1=\left[
\begin{smallmatrix}
K&0&0&0\\
0&-K&0&0\\
0&0&-K&0\\
0&0&0&K
\end{smallmatrix}\right],\
\hat{e}_2=\left[
\begin{smallmatrix}
0&K&0&0\\
K&0&0&0\\
0&0&0&K\\
0&0&K&0
\end{smallmatrix}\right],\
\hat{e}_3=\left[
\begin{smallmatrix}
0&-K&0&0\\
K&0&0&0\\
0&0&0&-K\\
0&0&K&0
\end{smallmatrix}\right],\
\hat{e}_4=\left[
\begin{smallmatrix}
0&0&-K&0\\
0&0&0&K\\
K&0&0&0\\
0&-K&0&0
\end{smallmatrix}\right],
\]
where $K$ is a shortcut for division ring element $K=\{K(1)\}$, which in  for \cl{2}{2} is represented by a single entry~$1$ in $4\times 4$
matrix $E_{ij}$. Thus, the matrix rep of GA basis elements (generators) in \cl{2}{2} is
\[\kern-2em
\hat{e}_1=\left[
\begin{smallmatrix}
1&0&0&0\\
0&-1&0&0\\
0&0&-1&0\\
0&0&0&1
\end{smallmatrix}\right],\
\hat{e}_2=\left[
\begin{smallmatrix}
0&1&0&0\\
1&0&0&0\\
0&0&0&1\\
0&0&1&0
\end{smallmatrix}\right],\
\hat{e}_3=\left[
\begin{smallmatrix}
0&-1&0&0\\
1&0&0&0\\
0&0&0&-1\\
0&0&1&0
\end{smallmatrix}\right],\
\hat{e}_4=\left[
\begin{smallmatrix}
0&0&-1&0\\
0&0&0&1\\
1&0&0&0\\
0&-1&0&0
\end{smallmatrix}\right].
\]

If the division ring $K$ contains more elements, then all such elements will appear in the calculated $\hat{e}_i$ matrices. In order to get
$\bbR$, $\bbC$, or $\bbH$ matrix reps we need to replace the division ring element by the corresponding isomorphic element $(1)$ or $(1,\ii)$,
or $(q_0=1,q_1,q_2,q_3)$, where $q_i$ denotes the quaternion components.

 If $\e{k}$ is replaced by elements of ideal basis $S_k$, namely
$E_{ij}(S_k)=\reverse{S}^iS_kS_j$, then one can obtain the matrices having only a single unity in the first column,
\[E_{11}=\left[
\begin{smallmatrix}1&0&0&0\\
0&0&0&0\\
0&0&0&0\\
0&0&0&0
\end{smallmatrix}\right],\quad
E_{21}=\left[
\begin{smallmatrix}0&0&0&0\\
1&0&0&0\\
0&0&0&0\\
0&0&0&0
\end{smallmatrix}\right],\quad
E_{31}=\left[
\begin{smallmatrix}0&0&0&0\\
0&0&0&0\\
1&0&0&0\\
0&0&0&0
\end{smallmatrix}\right],\quad
E_{41}=\left[
\begin{smallmatrix}0&0&0&0\\
0&0&0&0\\
0&0&0&0\\
1&0&0&0
\end{smallmatrix}\right], \]
what  indicates that the ideal basis elements make up a spinorial basis.

If the division ring contains more than one element, i.e., it is a complex field or a quaternion ring, then the column spinor can be obtained by
 sum $\Psi=\sum_{j,k,i} s_{jk} K_i(k) S_1(j)$, where $s_{jk}$ denotes an arbitrary real scalar and,  $K_i$ and $S_i$ is  the division ring and
the ideal basis  of the idempotent $P_i$ (see the Spinor data tables for explicit $P_i$ expression), respectively.

\item[5.] \textit{General spinor $\Psi$}. A general spinor is the sum of generators (item~3) multiplied by real
coefficients $s_i$,
\[\begin{split}
\Psi=&(s_1+s_2\e{2}+s_3\e{4}+s_4\e{24})P_1=\\
&\tfrac{1}{4}s_1(1+\e{1}+\e{23}+\e{123})+\tfrac{1}{4}s_2(\e{2}+\e{3}-\e{12}-\e{13})+\\
&\tfrac{1}{4}s_3(\e{4}-\e{14}+\e{234}-\e{1234})+\tfrac{1}{4}s_4(\e{24}+\e{34}+\e{124}+\e{134}).
\end{split}
\]

\item[6.] \textit{Spinor $\Psi$ in matrix form}. In the matrix form the general MV spinor can be expanded in the ideal basis as
\[\hat{\Psi}=s_1E_{11}+s_2E_{21}+s_3E_{31}+s_4E_{41}\,.
 \]

\item[7.] \textit{Square of spinor norm} is
  \[\norm{\Psi}^2=\langle\Psi^\dagger\Psi\rangle= \langle\reverse{\Psi}^{\sharp}\Psi\rangle=-\langle\e{34}\widetilde{\Psi}\e{34} \Psi\rangle=\frac14(s_1^2+s_2^2+s_3^2+s_4^2),
\]
where the tilde and brackets indicate, respectively, the reversion and  scalar part projection.

\end{itemize}
The results of the example are summarized in  Table~\ref{spt4} (\cl{2}{2} algebra). The implementation and step by step computations are  given in~\cite{AcusDargys2024} (the notebook {\it  10\_AlgebraicSpinorsOfRealCl.nb}), where data for GAs up to $n=8$ are explicitly computed and presented.

\subsection{Norm of spinor in real GA.}\label{spinorNorm} In spinor data Tables the MVs have standard form, i.e.,
they are written with lower indices. The norm  may be rewritten more uniformly in algebra independent way if upper (related to reciprocal basis) and lower indices are used simultaneously. Then, with the reciprocal orthonormalized frame introduced  the spinor norm is
$\norm{\Psi}^2=\langle\Psi^\dagger\Psi\rangle=\langle\reverse{\Psi}^{\sharp}\Psi\rangle_0$, where $\Psi$ is a general spinor and $\Psi^{\sharp}$ is the spinor with all indices
raised, i.e.  in the reciprocal basis (cf. item~4 in the Examples~\ref{CL22exampleB}). Then, the spinor $\Psi^{\sharp}$ with all upper indices
after index lowering gives a correct  sign.

 The normalization factor $2^k$ in the real spinor data Tables (last row, left-side of equation),
where $k$ counts a number of factors in the primitive idempotent $P_i$, ensures that the hermitian norm computed either by formula
$\norm{\Psi}^2= \langle\Psi^\dagger\Psi\rangle=\langle\reverse{\Psi}^{\sharp}\Psi\rangle$, or its matrix counterpart $\norm{\hat{\Psi}}^2=2^{-k} \Tr (\hat{\Psi}^\dagger
\hat{\Psi})$ coincides. In  real Euclidean spaces the norm is $\norm{\Psi}^2= \langle\reverse{\Psi}\Psi\rangle$,
since index raising/lowering operation has no effect upon sign of  basis elements~\cite{Hestenes66}.


\section{Complex GA spinors data tables (complex idempotent)\label{complexGAspinors}}
In real GAs, not all basis elements are acceptable for construction of primitive idempotents. Specially, the squares of elements are
required to be equal to $+1$. If multiplication of basis elements by imaginary unit is permitted, then any $\e{J}$ may be
used in idempotent construction. Complex spinor data Tables were constructed by allowing complex idempotents. Both the computation procedure and the notations in the Tables below remain the same. The asterisk $(*)$ denotes complex conjugation.

\subsection{Even complex algebras}

\begin{flushleft}
\begin{table}[H]
\centering
\begin{tabular}{l|>{$}l<{$}>{$}l<{$}>{$}l<{$}}
  &  \cl{2}{0}[\bbC(2)]& \cl{1}{1}[\bbC(2)] & \cl{0}{2}[\bbC(2)]
\\ \hline
1&
t=\tfrac12 (1 + \e{1})
&
t=\tfrac12 (1 + \e{1})
&
t=\tfrac12 (1 + \ii \e{1})
\\
2&
\{t1t\}=\tfrac12 (1 + \ii \e{1}) &
\{t1t\}=\tfrac12(1+\e{1}) &
\{t1t\}=\tfrac12 (1 + \ii\e{1})
\\
3&
\begin{multlined}[t][0.1\columnwidth]
\{t,\e{2}t\}
\end{multlined}
&
\begin{multlined}[t][0.1\columnwidth]
\{t,\e{2}t\}
\end{multlined}
&
\begin{multlined}[t][0.1\columnwidth]
\{t,\e{2}t\}
\end{multlined}
\\
4&
\begin{multlined}[t][0.1\columnwidth]
\{E_{11} - E_{22}, E_{12} + E_{21}\}
\end{multlined}
&
\begin{multlined}[t][0.1\columnwidth]\{E_{11} - E_{22}, -E_{12} + E_{21}\}
\end{multlined}
&
\begin{multlined}[t][0.1\columnwidth]
  \{\ii(-E_{11} + E_{22}), -E_{12} + E_{21}\}
\end{multlined}
\\
5&
\begin{multlined}[t][0.1\columnwidth]
\Psi=(s_{1} + s_{2}\e{2})t
\end{multlined}
&
\begin{multlined}[t][0.1\columnwidth]
\Psi=(s_{1} + s_{2}\e{2})t
\end{multlined}
&
\begin{multlined}[t][0.1\columnwidth]
  \Psi=(s_{1}+s_{2}\e{2})t
\end{multlined}
\\
6&
s_{1}E_{11} +s_{2} E_{21}
&
s_{1}E_{11} +s_{2} E_{21}
&
\begin{multlined}[t][0.1\columnwidth]
s_{1}E_{11} +s_{2} E_{21}
\end{multlined}
\\
7&
\begin{multlined}[t][0.1\columnwidth]
\langle\reverse{\Psi}\Psi\rangle
=\tfrac12(|s_1|^2+|s_2|^2)
\end{multlined}
&
\begin{multlined}[t][0.1\columnwidth]
\langle\e{1}\reverse{\Psi}\e{1}\Psi\rangle
=\tfrac12(|s_1|^2+|s_2|^2)
\end{multlined}
&
\begin{multlined}[t][0.1\columnwidth]
  \langle \cliffordconjugate{\Psi}^*\Psi\rangle
=\tfrac12(|s_1|^2+|s_2|^2)
\end{multlined}
\end{tabular}
\caption{Complex spinor tables for GAs with vector space dimension $n=2$}\label{sptc1}
\end{table}

\begin{table}[H]
\centering
\begin{tabular}{l|>{$}l<{$}>{$}l<{$}>{$}l<{$}}
& \cl{4}{0}[{\bbC(4)}]& \cl{3}{1}[\bbC(4)]
\\ \hline
1&
t=\tfrac14 (1 + \e{1})(1 + \ii \e{23}) &
t=\tfrac14 (1 + \e{1})(1 + \ii \e{23})
\\
2&
\begin{multlined}[t][0.1\columnwidth]
  \{t 1 t \}
\end{multlined}
&
\begin{multlined}[t][0.1\columnwidth]
  \{t 1 t \}
\end{multlined}
\\
3&
\begin{multlined}[t][0.1\columnwidth]
\{t,\ii \e{2} t, \ii \e{4} t, \e{24} t\}
\end{multlined}
&
\begin{multlined}[t][0.1\columnwidth]
\{t,\ii \e{2} t, \ii \e{4} t, \e{24} t\}
\end{multlined}
\\
4&
\begin{multlined}[t][0.1\columnwidth]
\{E_{11} - E_{22} - E_{33} + E_{44},
  \ii(E_{12} - E_{21} - E_{34} + E_{43}),\\[-3ex]
 -E_{12} - E_{21} + E_{34} + E_{43},
 \ii(E_{13} + E_{24} - E_{31} - E_{42})
\}
\end{multlined}
&
\begin{multlined}[t][0.1\columnwidth]
\{E_{11} - E_{22} - E_{33} + E_{44},
  \ii(E_{12} - E_{21} - E_{34} + E_{43}),\\[-3ex]
 -E_{12} - E_{21} + E_{34} + E_{43},
 -\ii(E_{13} + E_{24} + E_{31} + E_{42})
\}
\end{multlined}
\\
5&
\begin{multlined}[t][0.1\columnwidth]
\Psi=(s_{1} + \ii s_{2}\e{2} + \ii s_{3}\e{4} + s_{4}\e{24}) t
\end{multlined}
&
\begin{multlined}[t][0.1\columnwidth]
\Psi=(s_{1} + \ii s_{2}\e{2} + \ii s_{3}\e{4} + s_{4}\e{24}) t
\end{multlined}
\\
6&
s_{1}E_{11} +s_{2} E_{21}+s_{3} E_{31}+s_{4} E_{41}
&
s_{1}E_{11} +s_{2} E_{21}+s_{3} E_{31}+s_{4} E_{41}
\\
7&
\langle\reverse{\Psi}^*\Psi\rangle=\frac14(|s_1|^2+\cdots+|s_4|^2)
&
-\langle\e{4}\cliffordconjugate{\Psi}^*\e{4}\Psi\rangle=\tfrac14(|s_1|^2+\cdots+|s_4|^2)
\end{tabular}
\vskip 10 pt

\begin{tabular}{l|>{$}l<{$}>{$}l<{$}>{$}l<{$}}
& \cl{2}{2}[{\bbC(4)}]& \cl{1}{3}[\bbC(4)]
\\ \hline
1&
t=\tfrac14 (1 + \e{1})(1 + \e{23}) &
t=\tfrac14 (1 + \e{1})(1 + \ii \e{23})
\\
2&
\begin{multlined}[t][0.1\columnwidth]
  \{t 1 t \}
\end{multlined}
&
\begin{multlined}[t][0.1\columnwidth]
  \{t 1 t \}
\end{multlined}
\\
3&
\begin{multlined}[t][0.1\columnwidth]
\{t,\e{2} t, \e{4} t, \e{24} t\}
\end{multlined}
&
\begin{multlined}[t][0.1\columnwidth]
\{t,\ii \e{2} t, \ii \e{4} t, \e{24} t\}
\end{multlined}
\\
4&
\begin{multlined}[t][0.1\columnwidth]
\{E_{11} - E_{22} - E_{33} + E_{44},
  E_{12} + E_{21} + E_{34} + E_{43},\\[-3ex]
 -E_{12} + E_{21} - E_{34} + E_{43},
 -E_{13} + E_{24} + E_{31} - E_{42}
\}
\end{multlined}
&
\begin{multlined}[t][0.1\columnwidth]
\{E_{11} - E_{22} - E_{33} + E_{44},
  \ii(-E_{12} - E_{21} + E_{34} + E_{43}),\\[-3ex]
 -E_{12} + E_{21} + E_{34} - E_{43},
 -\ii(E_{13} + E_{24} + E_{31} + E_{42})
\}
\end{multlined}
\\
5&
\begin{multlined}[t][0.1\columnwidth]
\Psi=(s_{1} + s_{2}\e{2} + s_{3}\e{4} + s_{4}\e{24}) t
\end{multlined}
&
\begin{multlined}[t][0.1\columnwidth]
\Psi=(s_{1} + \ii s_{2}\e{2} + \ii s_{3}\e{4} + s_{4}\e{24}) t
\end{multlined}
\\
6&
s_{1}E_{11} +s_{2} E_{21}+s_{3} E_{31}+s_{4} E_{41}
&
s_{1}E_{11} +s_{2} E_{21}+s_{3} E_{31}+s_{4} E_{41}
\\
7&
\begin{multlined}[t][0.1\columnwidth]
-\langle\e{34}\reverse{\Psi}^*\e{34}\Psi\rangle=\tfrac14(|s_1|^2+\cdots+|s_4|^2)
\end{multlined}
&
\begin{multlined}[t][0.1\columnwidth]
\langle\e{1}\widetilde{\Psi}^{*}\e{1}\Psi\rangle=\tfrac14(|s_1|^2+\cdots+|s_4|^2)
\end{multlined}
\end{tabular}
\vskip 10pt

\begin{tabular}{l|>{$}l<{$}}
& \cl{0}{4}[{\bbC(4)}]
\\ \hline
1&
t=\tfrac14 (1 + \ii \e{1})(1 + \ii \e{23})
\\
2&
\begin{multlined}[t][0.1\columnwidth]
  \{t 1 t \}
\end{multlined}
\\
3&
\begin{multlined}[t][0.1\columnwidth]
\{t,\e{2} t, \ii \e{4} t, \ii \e{24} t\}
\end{multlined}
\\
4&
\begin{multlined}[t][0.1\columnwidth]
  \{\ii (-E_{11} + E_{22} + E_{33} - E_{44}),
  -E_{12} + E_{21} - E_{34} + E_{43},
  \ii (E_{12} + E_{21} + E_{34} + E_{43}),
  \ii (-E_{13} + E_{24} - E_{31} + E_{42})
\}
\end{multlined}
\\
5&
\begin{multlined}[t][0.1\columnwidth]
\Psi=(s_{1} + s_{2}\e{2} + \ii s_{3}\e{4} + \ii s_{4}\e{24}) t
\end{multlined}
\\
6&
s_{1}E_{11} +s_{2} E_{21}+s_{3} E_{31}+s_{4} E_{41}
\\
7&
\langle\cliffordconjugate{\Psi}^*\Psi\rangle=\tfrac14(|s_1|^2+\cdots+|s_4|^2)
\end{tabular}
\caption{Complex spinor tables for GAs with vector space dimension $n=4$}\label{sptc2}
\end{table}
\vskip 10pt

\begin{table}[H]
\centering
\begin{tabular}{l|>{$}l<{$}}
& \cl{6}{0}[\bbC(8)]
\\ \hline
1&
t=\tfrac18 (1 + \e{1})(1 + \ii \e{23})(1 + \ii \e{45})
\\
2&
\begin{multlined}[t][0.1\columnwidth]
  \{t 1 t \}
\end{multlined}
\\
3&
\begin{multlined}[t][0.1\columnwidth]
\{t,\e{2} t, \e{4} t, \e{24} t,\e{6} t,\e{26} t,\e{46} t,\e{246}
t\}
\end{multlined}
\\
4&
\begin{multlined}[t][0.1\columnwidth]
\{E_{11} - E_{22} - E_{33} + E_{44} - E_{55} + E_{66} + E_{77} - E_{88},
E_{12} + E_{21} + E_{34} + E_{43} + E_{56} + E_{65} + E_{78} + E_{87},\\[-3ex]
\ii(E_{12} - E_{21} + E_{34} - E_{43} + E_{56} - E_{65} + E_{78} - E_{87}),
E_{13} - E_{24} + E_{31} - E_{42} + E_{57} - E_{68} + E_{75} - E_{86},\\[-1ex]
\ii(E_{13} - E_{24} - E_{31} + E_{42} + E_{57} - E_{68} - E_{75} + E_{86}),
E_{15} - E_{26} - E_{37} + E_{48} + E_{51} - E_{62} - E_{73} + E_{84}
\}
\end{multlined}
\\
5&
\begin{multlined}[t][0.1\columnwidth]
 \Psi= (s_{1} + s_{2}\e{2} + s_{3}\e{4} + s_{4}\e{24} + s_{5}\e{6}
 + s_{6}\e{26} + s_{7}\e{46} + s_{8}\e{246}) t
\end{multlined}
\\
6&
s_{1}E_{11} +s_{2} E_{21}+s_{3} E_{31}+s_{4} E_{41}+s_{5}E_{51} +s_{6} E_{61}+s_{7} E_{71}+s_{8} E_{81}
\\
7&
\langle\reverse{\Psi}^*\Psi\rangle=\tfrac18(|s_1|^2+\cdots+|s_8|^2)
\end{tabular}
\vskip 10pt

\begin{tabular}{l|>{$}l<{$}}
& \cl{5}{1}[\bbC(8)]
\\ \hline
1&
t=\tfrac18 (1 + \e{1})(1 + \ii \e{23})(1 + \ii \e{45})
\\
2&
\begin{multlined}[t][0.1\columnwidth]
  \{t 1 t \}
\end{multlined}
\\
3&
\begin{multlined}[t][0.1\columnwidth]
\{t,\e{2} t, \e{4} t, \e{24} t,\e{6} t,\e{26} t,\e{46} t,\e{246}
t\}
\end{multlined}
\\
4&
\begin{multlined}[t][0.1\columnwidth]
\{E_{11} - E_{22} - E_{33} + E_{44} - E_{55} + E_{66} + E_{77} - E_{88},
E_{12} + E_{21} + E_{34} + E_{43} + E_{56} + E_{65} + E_{78} + E_{87},\\[-3ex]
\ii(E_{12} - E_{21} + E_{34} - E_{43} + E_{56} - E_{65} + E_{78} - E_{87}),
E_{13} - E_{24} + E_{31} - E_{42} + E_{57} - E_{68} + E_{75} - E_{86},\\[-1ex]
\ii(E_{13} - E_{24} - E_{31} + E_{42} + E_{57} - E_{68} - E_{75} + E_{86}),
-E_{15} + E_{26} + E_{37} - E_{48} + E_{51} - E_{62} - E_{73} + E_{84}
\}
\end{multlined}
\\
5&
\begin{multlined}[t][0.1\columnwidth]
\Psi=(s_{1} + s_{2}\e{2} + s_{3}\e{4} + s_{4}\e{24} + s_{5}\e{6} +
s_{6}\e{26} + s_{7}\e{46} + s_{8}\e{246}) t
\end{multlined}
\\
6&
s_{1}E_{11} +s_{2} E_{21}+s_{3} E_{31}+s_{4} E_{41}+s_{5}E_{51} +s_{6} E_{61}+s_{7} E_{71}+s_{8} E_{81}
\\
7&
-\langle\e{6}\cliffordconjugate{\Psi}^*\e{6}\Psi\rangle=\tfrac18(|s_1|^2+\cdots+|s_8|^2)
\end{tabular}
\vskip 10 pt

\begin{tabular}{l|>{$}l<{$}}
& \cl{4}{2}[\bbC(8)]
\\ \hline
1&
t=\tfrac18 (1 + \e{1})(1 + \ii \e{23})(1 + \e{45})
\\
2&
\begin{multlined}[t][0.1\columnwidth]
  \{t 1 t\}
\end{multlined}
\\
3&
\begin{multlined}[t][0.1\columnwidth]
\{t,\ii \e{2} t, \ii \e{4} t, \e{24} t,\ii \e{6} t,\e{26} t,\e{46}
t,\ii \e{246} t\}
\end{multlined}
\\
4&
\begin{multlined}[t][0.1\columnwidth]
\{E_{11} - E_{22} - E_{33} + E_{44} - E_{55} + E_{66} + E_{77} - E_{88},
\ii( E_{12} - E_{21} - E_{34} + E_{43} - E_{56} + E_{65} + E_{78} - E_{87}),\\[-3ex]
-E_{12} - E_{21} + E_{34} + E_{43} + E_{56} + E_{65} - E_{78} - E_{87},
\ii(E_{13} + E_{24} - E_{31} - E_{42} - E_{57} - E_{68} + E_{75} + E_{86}),\\[-1ex]
\ii(-E_{13} - E_{24} - E_{31} - E_{42} + E_{57} + E_{68} + E_{75} + E_{86}),
\ii(-E_{15} - E_{26} - E_{37} - E_{48} - E_{51} - E_{62} - E_{73} - E_{84})
\}
\end{multlined}
\\
5&
\begin{multlined}[t][0.1\columnwidth]
\Psi=(s_{1} + \ii s_{2}\e{2} + \ii s_{3}\e{4} + s_{4}\e{24} + \ii
s_{5}\e{6} + s_{6}\e{26} + s_{7}\e{46} + \ii s_{8}\e{246}) t
\end{multlined}
\\
6&
s_{1}E_{11} +s_{2} E_{21}+s_{3} E_{31}+s_{4} E_{41}+s_{5}E_{51} +s_{6} E_{61}+s_{7} E_{71}+s_{8} E_{81}
\\
7&
-\langle\e{56}\reverse{\Psi}^*\e{56}\Psi\rangle=\tfrac18(|s_1|^2+\cdots+|s_8|^2)
\end{tabular}
\vskip 10pt

\begin{tabular}{l|>{$}l<{$}}
& \cl{3}{3}[\bbC(8)]\index{algebra!\cl{3}{3}}
\\ \hline
1&
t=\tfrac18 (1 + \e{1})(1 + \ii \e{23})(1 + \ii\e{45})
\\
2&
\begin{multlined}[t][0.1\columnwidth]
  \{t 1 t \}
\end{multlined}
\\
3&
\begin{multlined}[t][0.1\columnwidth]
\{t,\e{2} t, \e{4} t, \ii \e{24} t, \e{6} t,\e{26} t, \ii \e{46}
t, \e{246} t\}
\end{multlined}
\\
4&
\begin{multlined}[t][0.1\columnwidth]
\{E_{11} - E_{22} - E_{33} + E_{44} - E_{55} + E_{66} + E_{77} - E_{88},
E_{12} + E_{21} + \ii E_{34} - \ii E_{43} + E_{56} + E_{65} - \ii E_{78} + \ii E_{87},\\[-3ex]
\ii E_{12} - \ii E_{21} - E_{34} - E_{43} + \ii E_{56} - \ii E_{65} + E_{78} + E_{87},
 -E_{13} + \ii E_{24} + E_{31} + \ii E_{42} - \ii E_{57} + E_{68} - \ii E_{75} - E_{86},\\[-1ex]
\ii E_{13} + E_{24} + \ii E_{31} - E_{42} - E_{57} - \ii E_{68} + E_{75} - \ii E_{86},
-E_{15} + E_{26} + \ii E_{37} + \ii E_{48} + E_{51} - E_{62} + \ii E_{73} + \ii E_{84}
\}
\end{multlined}
\\
5&
\begin{multlined}[t][0.1\columnwidth]
\Psi=(s_{1} + s_{2}\e{2} + s_{3}\e{4}  + \ii s_{4}\e{24}  +
s_{5}\e{6} + s_{6}\e{26} + \ii s_{7}\e{46} + s_{8}\e{246}) t
\end{multlined}
\\
6&
s_{1}E_{11} +s_{2} E_{21}+s_{3} E_{31}+s_{4} E_{41}+s_{5}E_{51} +s_{6} E_{61}+s_{7} E_{71}+s_{8} E_{81}
\\
7&
-\langle\e{123}\reverse{\Psi}^*\e{123}\Psi\rangle=
\tfrac18(|s_1|^2+\cdots+|s_8|^2)
\end{tabular}
\vskip 10pt

\begin{tabular}{l|>{$}l<{$}}
& \cl{2}{4}[\bbC(8)]
\\ \hline
1&
t=\tfrac18 (1 + \e{1})(1 + \e{23})(1 + \ii\e{45})
\\
2&
\begin{multlined}[t][0.1\columnwidth]
  \{t 1 t \}
\end{multlined}
\\
3&
\begin{multlined}[t][0.1\columnwidth]
\{t,\ii \e{2} t, \ii \e{4} t, \e{24} t, \ii \e{6} t,\e{26} t,
\e{46} t, \ii \e{246} t\}
\end{multlined}
\\
4&
\begin{multlined}[t][0.1\columnwidth]
\{E_{11} - E_{22} - E_{33} + E_{44} - E_{55} + E_{66} + E_{77} - E_{88},
  \ii(E_{12} - E_{21} - E_{34} + E_{43} - E_{56} + E_{65} + E_{78} - E_{87}),\\[-3ex]
  \ii(-E_{12} - E_{21} + E_{34} + E_{43} + E_{56} + E_{65} - E_{78} - E_{87}),
  \ii(-E_{13} - E_{24} - E_{31} - E_{42} + E_{57} + E_{68} + E_{75} + E_{86}),\\[-1ex]
-E_{13} - E_{24} + E_{31} + E_{42} + E_{57} + E_{68} - E_{75} - E_{86},
\ii(-E_{15} - E_{26} - E_{37} - E_{48} - E_{51} - E_{62} - E_{73} - E_{84})
\}
\end{multlined}
\\
5&
\begin{multlined}[t][0.1\columnwidth]
\Psi=(s_{1} + \ii s_{2}\e{2} + \ii s_{3}\e{4} + s_{4}\e{24} + \ii
s_{5}\e{6} + s_{6}\e{26} + s_{7}\e{46} + \ii s_{8}\e{246}) t
\end{multlined}
\\
6&
s_{1}E_{11} +s_{2} E_{21}+s_{3} E_{31}+s_{4} E_{41}+s_{5}E_{51} +s_{6} E_{61}+s_{7} E_{71}+s_{8} E_{81}
\\
7&
-\langle\e{12}\cliffordconjugate{\Psi}^*\e{12}\Psi\rangle=
\tfrac18(|s_1|^2+\cdots+|s_8|^2)
\end{tabular}
\vskip 10pt

\begin{tabular}{l|>{$}l<{$}}
& \cl{1}{5}[\bbC(8)]
\\ \hline
1&
t=\tfrac18 (1 + \e{1})(1 + \ii \e{23})(1 + \ii\e{45})
\\
2&
\begin{multlined}[t][0.1\columnwidth]
  \{t 1 t \}
\end{multlined}
\\
3&
\begin{multlined}[t][0.1\columnwidth]
\{t,\e{2} t, \e{4} t, \e{24} t, \e{6} t, \ii \e{26} t, \ii \e{46}
t, \e{246} t\}
\end{multlined}
\\
4&
\begin{multlined}[t][0.1\columnwidth]
\{E_{11} - E_{22} - E_{33} + E_{44} - E_{55} + E_{66} + E_{77} - E_{88},
   -E_{12} + E_{21} - E_{34} + E_{43} - \ii E_{56} - \ii E_{65} + \ii E_{78} + \ii E_{87},\\[-3ex]
  \ii E_{12} + \ii E_{21} + \ii E_{34} + \ii E_{43} - E_{56} + E_{65} + E_{78} - E_{87},
   -E_{13} + E_{24} + E_{31} - E_{42} - \ii E_{57} - \ii E_{68} - \ii E_{75} - \ii E_{86},\\[-1ex]
\ii E_{13} - \ii E_{24} + \ii E_{31} - \ii E_{42} - E_{57} - E_{68} + E_{75} + E_{86},
-E_{15} + \ii E_{26} + \ii E_{37} - E_{48} + E_{51} + \ii E_{62} + \ii E_{73} + E_{84}
\}
\end{multlined}
\\
5&
\begin{multlined}[t][0.1\columnwidth]
\Psi=(s_{1} + s_{2}\e{2} + s_{3}\e{4}  + s_{4}\e{24} + s_{5}\e{6}
+ \ii s_{6}\e{26} + \ii s_{7}\e{46} + s_{8}\e{246}) t
\end{multlined}
\\
6&
s_{1}E_{11} +s_{2} E_{21}+s_{3} E_{31}+s_{4} E_{41}+s_{5}E_{51} +s_{6} E_{61}+s_{7} E_{71}+s_{8} E_{81}
\\
7&
\langle\e{1}\reverse{\Psi}^*\e{1}\Psi\rangle=
\tfrac18(|s_1|^2+\cdots+|s_8|^2)
\end{tabular}
\caption{Complex spinor tables for $\cl{6}{0}$, $\cl{5}{1}$, $\cl{4}{2}$, $\cl{3}{3}$, $\cl{2}{4}$ and $\cl{1}{5}$; $n=6$}\label{sptc3a}
\end{table}

\begin{table}[H]
\centering
\begin{tabular}{l|>{$}l<{$}}
& \cl{0}{6}[\bbC(8)]
\\ \hline
1&
t=\tfrac18 (1 + \ii \e{1})(1 + \ii \e{23})(1 + \ii\e{45})
\\
2&
\begin{multlined}[t][0.1\columnwidth]
  \{t 1 t \}
\end{multlined}
\\
3&
\begin{multlined}[t][0.1\columnwidth]
\{\ii t, \ii \e{2} t, \ii \e{4} t, \ii \e{24} t, \ii \e{6} t, \ii
\e{26} t, \ii \e{46} t, \ii \e{246} t\}
\end{multlined}
\\
4&
\begin{multlined}[t][0.1\columnwidth]
  \{\ii (-E_{11} + E_{22} + E_{33} - E_{44} + E_{55} - E_{66} - E_{77} + E_{88}),
   -E_{12} + E_{21} - E_{34} + E_{43} - E_{56} + E_{65} - E_{78} + E_{87},\\[-3ex]
   \ii (E_{12} + E_{21} + E_{34} + E_{43} + E_{56} + E_{65} + E_{78} + E_{87}),
  -E_{13} + E_{24} + E_{31} - E_{42} - E_{57} + E_{68} + E_{75} - E_{86},\\[-1ex]
  \ii (E_{13} - E_{24} + E_{31} - E_{42} + E_{57} - E_{68} + E_{75} - E_{86}),
-E_{15} + E_{26} + E_{37} - E_{48} + E_{51} - E_{62} - E_{73} + E_{84}
\}
\end{multlined}
\\
5&
\begin{multlined}[t][0.1\columnwidth]
\Psi=\ii (s_{1} + s_{2}\e{2} + s_{3}\e{4}+ s_{4}\e{24} +
s_{5}\e{6} + s_{6}\e{26} + s_{7}\e{46} + s_{8}\e{246}) t
\end{multlined}
\\
6&
\ii (s_{1}E_{11} +s_{2} E_{21}+s_{3} E_{31}+s_{4} E_{41}+s_{5}E_{51} +s_{6} E_{61}+s_{7} E_{71}+s_{8} E_{81})
\\
7&
\langle\cliffordconjugate{\Psi}^*\Psi\rangle=\tfrac18(|s_1|^2+\cdots+|s_8|^2)
\end{tabular}
\caption{Complex spinor tables for $\cl{0}{6}$; n=6}\label{sptc3b}
\end{table}

\end{flushleft}
\subsection{Odd complex algebras}

In matrix rep, the odd complex algebras consist of two blocks on a diagonal. The  diagonal elements in the first and second block come from,
respectively, left basis ideal sorted in `RevLex' order and after application of grade inversion to left ideal basis, which then is sorted
in`InvLex' order. Such a procedure ensures
that in spinor matrix  rep the first nonzero column of spinor always appears as a first column, and the second nonzero column appears in the
middle of the matrix, i.e., the spinor matches the block diagonal structure of MV matrix rep). The coefficients $s_i$ now are complex numbers.  The block diagonal structure reflects the fact that
the underlying field $\{1,\ii\}$ is commutative. Contrary to the case of real algebras, the block diagonal matrices, $\bbC(n)\oplus\bbC(n)$, of
the complex spinor have $n$ independent coefficients instead of~$2n$. Let us remind that in a semi-simple real algebra (i.e. when $p-q=1\
\textrm{mod}\ 4$), the spinors have $2n$ free coefficients.

\begin{flushleft}
\begin{table}[H]
\centering
\begin{tabular}{l|>{$}l<{$}>{$}l<{$}>{$}l<{$}}
& \cl{3}{0}[{\bbC(2)\oplus\bbC(2)}]&
\cl{2}{1}[\bbC(2)\oplus\bbC(2)]
\\ \hline
1&
t=\tfrac12 (1 + \e{1}) &
t=\tfrac12 (1 + \e{1})
\\
2&
\begin{multlined}[t][0.1\columnwidth]
  \{t 1 t, t \e{23} t \}
\end{multlined}
&
\begin{multlined}[t][0.1\columnwidth]
  \{t 1 t, t \ii \e{23} t \}
\end{multlined}
\\
3&
\begin{multlined}[t][0.1\columnwidth]
\{t, \e{2} t \}
\end{multlined}
&
\begin{multlined}[t][0.1\columnwidth]
\{t, \e{2} t \}
\end{multlined}
\\
4&
\begin{multlined}[t][0.1\columnwidth]
\{E_{11} - E_{22} + E_{33} - E_{44},
 E_{12} + E_{21} - E_{34} - E_{43},\\[-3ex]
 \ii(-E_{12} + E_{21} - E_{34} + E_{43})
\}
\end{multlined}
&
\begin{multlined}[t][0.1\columnwidth]
\{E_{11} - E_{22} + E_{33} - E_{44},
 E_{12} + E_{21} - E_{34} - E_{43},\\[-3ex]
 -E_{12} + E_{21} - E_{34} + E_{43}
\}
\end{multlined}
\\
5&
\begin{multlined}[t][0.1\columnwidth]
\Psi=(s_{1} + s_{2}\e{23} + s_{3}\e{2} - s_{4}\e{3}) t
\end{multlined}
&
\begin{multlined}[t][0.1\columnwidth]
\Psi=(s_{1} + \ii s_{2}\e{23} + s_{3}\e{2} - \ii s_{4}\e{3}) t
\end{multlined}
\\
6&
\begin{multlined}[t][0.1\columnwidth]
(s_{1} + \ii s_{2})E_{11} +(s_{3} - \ii s_{4}) E_{21} \\[-3ex]+(s_{1} - \ii s_{2})E_{33}  + (-s_{3} - \ii s_{4})E_{43}
\end{multlined}
&
\begin{multlined}[t][0.1\columnwidth]
(s_{1} + \ii s_{2})E_{11} +(s_{3} - \ii s_{4}) E_{21} \\[-3ex]+(s_{1} - \ii s_{2})E_{33}  + (-s_{3} - \ii s_{4})E_{43}
\end{multlined}
\\
7&
\begin{multlined}[t][0.1\columnwidth]
\langle\reverse{\Psi}^*\Psi\rangle=\tfrac12(|s_1|^2+\cdots+|s_4|^2)
\end{multlined}
&
\begin{multlined}[t][0.1\columnwidth]
-\langle\e{3}\cliffordconjugate{\Psi}^*\e{3}\Psi\rangle=\tfrac12(|s_1|^2+\cdots+|s_4|^2)
\end{multlined}
\end{tabular}
\vskip 10pt

\begin{tabular}{l|>{$}l<{$}>{$}l<{$}>{$}l<{$}}
& \cl{1}{2}[{\bbC(2)\oplus\bbC(2)}]&
\cl{0}{3}[\bbC(2)\oplus\bbC(2)]
\\ \hline
1&
t=\tfrac12 (1 + \e{1}) &
t=\tfrac12 (1 + \ii \e{1})
\\
2&
\begin{multlined}[t][0.1\columnwidth]
  \{t 1 t, t \e{23} t \}
\end{multlined}
&
\begin{multlined}[t][0.1\columnwidth]
  \{t 1 t, t \e{23} t \}
\end{multlined}
\\
3&
\{t, \e{2} t \} &
\{t, \e{2} t \}
\\
4&
\begin{multlined}[t][0.1\columnwidth]
\{E_{11} - E_{22} + E_{33} - E_{44},
 -E_{12} + E_{21} - E_{34} + E_{43},\\[-3ex]
 \ii(-E_{12} - E_{21} + E_{34} + E_{43})
\}
\end{multlined}&
\begin{multlined}[t][0.1\columnwidth]
  \{\ii(-E_{11} + E_{22} - E_{33} + E_{44}),
 -E_{12} + E_{21} - E_{34} + E_{43},\\[-3ex]
 \ii(-E_{12} - E_{21} + E_{34} + E_{43})
\}
\end{multlined}
\\
5&
\begin{multlined}[t][0.1\columnwidth]
\Psi=(s_{1} + s_{2}\e{23} + s_{3}\e{2} + s_{4}\e{3}) t
\end{multlined}
&
\begin{multlined}[t][0.1\columnwidth]
\Psi=(s_{1} + s_{2}\e{23} + s_{3}\e{2} + s_{4}\e{3}) t
\end{multlined}
\\
6&
\begin{multlined}[t][0.1\columnwidth]
(s_{1} + \ii s_{2})E_{11}  + (s_{3} - \ii s_{4})E_{21}  \\[-3ex] + (s_{1} - \ii s_{2})E_{33}  + (s_{3} + \ii s_{4})E_{43}
\end{multlined}
&
\begin{multlined}[t][0.1\columnwidth]
(s_{1} + \ii s_{2})E_{11}  + (s_{3} - \ii s_{4})E_{21}  \\[-3ex] + (s_{1} - \ii s_{2})E_{33}  + (s_{3} + \ii s_{4})E_{43}
\end{multlined}
\\
7&
\begin{multlined}[t][0.1\columnwidth]
\langle\e{1}\reverse{\Psi}^*\e{1}\Psi\rangle=\tfrac12(|s_1|^2+\cdots+|s_4|^2)
\end{multlined}
&
\begin{multlined}[t][0.1\columnwidth]
\langle\cliffordconjugate{\Psi}^*\Psi\rangle=\tfrac12(|s_1|^2+\cdots+|s_4|^2)
\end{multlined}
\end{tabular}
\caption{Complex spinor tables for geometric algebras with vector space dimension $n=3$}\label{sptc4}
\end{table}
\vskip 5pt

\begin{table}[H]
\centering
\begin{tabular}{l|>{$}l<{$}>{$}l<{$}>{$}l<{$}}
& \cl{5}{0}[{\bbC(4)\oplus\bbC(4)}]&
\cl{4}{1}[{\bbC(4)\oplus\bbC(4)}]
\index{algebra!\cl{5}{0}}\index{algebra!\cl{4}{1}}
\\ \hline
1&
t=\tfrac14 (1 + \e{1})(1 + \ii \e{23})
&
t=\tfrac14 (1 + \e{1})(1 + \ii \e{23})
\\
2&
\begin{multlined}[t][0.1\columnwidth]
  \{t 1 t, t \e{45} t \}
\end{multlined}
&
\begin{multlined}[t][0.1\columnwidth]
  \{t 1 t, t (\ii \e{45}) t \}
\end{multlined}
\\
3&
\begin{multlined}[t][0.1\columnwidth]
\{t, \ii \e{2} t, \ii \e{4} t, \e{24} t \}
\end{multlined}
&
\begin{multlined}[t][0.1\columnwidth]
\{t, \ii \e{2} t, \ii \e{4} t, \e{24} t \}
\end{multlined}
\\
4&
\begin{multlined}[t][0.1\columnwidth]
  \{E_{11} - E_{22} - E_{33} + E_{44} + E_{55} - E_{66} - E_{77} + E_{88},\\[-3ex]
    \ii(E_{12} - E_{21} - E_{34} + E_{43} + E_{56} - E_{65} - E_{78} + E_{87}),
\\[-1ex]
-E_{12} - E_{21} + E_{34} + E_{43} - E_{56} - E_{65} + E_{78} + E_{87},
\\[-1ex]
\ii(E_{13} + E_{24} - E_{31} - E_{42} + E_{57} + E_{68} - E_{75} - E_{86}),
\\[-1ex] E_{13} + E_{24} + E_{31} + E_{42} - E_{57} - E_{68} - E_{75} - E_{86}
\}
\end{multlined}
&
\begin{multlined}[t][0.1\columnwidth]
\{
  E_{11} - E_{22} - E_{33} + E_{44} + E_{55} - E_{66} - E_{77} + E_{88},\\[-3ex]
  \ii(E_{12} - E_{21} - E_{34} + E_{43} + E_{56} - E_{65} - E_{78} + E_{87}),\\[-1ex]
  -E_{12} - E_{21} + E_{34} + E_{43}  - E_{56} - E_{65} + E_{78} + E_{87},\\[-1ex]
  \ii (E_{13} + E_{24} - E_{31} - E_{42}
  + E_{57} + E_{68} - E_{75} - E_{86}),\\[-1ex]
  \ii(-E_{13} - E_{24} - E_{31} - E_{42} + E_{57} + E_{68} + E_{75} + E_{86})
\}
\end{multlined}
\\
5&
\begin{multlined}[t][0.1\columnwidth]
\Psi=(s_{1} + s_{2}\e{45} + \ii s_{3}\e{2} + \ii s_{4}\e{245} + \ii s_{5}\e{4} - \ii s_{6}\e{5}\\[-3ex] + s_{7}\e{24} - s_{8}\e{25}) t
\end{multlined}
&
\begin{multlined}[t][0.1\columnwidth]
\Psi=(s_{1}  + \ii s_{2}\e{45}  + \ii s_{3}\e{2}- s_{4}\e{245}  + \ii s_{5}\e{4} + s_{6}\e{5}\\[-3ex] + s_{7}\e{24} - \ii s_{8}\e{25}) t
\end{multlined}
\\
6&
\begin{multlined}[t][0.1\columnwidth]
(s_{1} + \ii s_{2})E_{11} + (s_{3} + \ii s_{4})E_{21} + (s_{5} - \ii s_{6})E_{31}\\[-3ex]  + (s_{7} - \ii s_{8})E_{41} +(s_{1} - \ii s_{2})E_{55}  + (s_{3} - \ii s_{4})E_{65}\\[-1ex]   + (s_{5} + \ii s_{6})E_{75}  + (s_{7} + \ii s_{8})E_{85}
\end{multlined}
&
\begin{multlined}[t][0.1\columnwidth]
(s_{1} + \ii s_{2})E_{11} + (s_{3} + \ii s_{4})E_{21} +(s_{5} - \ii s_{6}) E_{31}\\[-3ex] + (s_{7} - \ii s_{8})E_{41} + (s_{1} - \ii s_{2})E_{55}  + (s_{3} - \ii s_{4})E_{65}\\[-1ex] + (s_{5} + \ii s_{6})E_{75}  + (s_{7} + \ii s_{8})E_{85}
\end{multlined}
\\
7&
\begin{multlined}[t][0.1\columnwidth]
\langle\reverse{\Psi}^*\Psi\rangle
=\tfrac14(|s_1|^2+\cdots+|s_8|^2)
\end{multlined}
&
\begin{multlined}[t][0.1\columnwidth]
  -\langle\e{5}\cliffordconjugate{\Psi}^*\e{5}\Psi\rangle
=\tfrac14(|s_1|^2+\cdots+|s_8|^2)
\end{multlined}
\end{tabular}
\vskip 10pt

\begin{tabular}{l|>{$}l<{$}>{$}l<{$}>{$}l<{$}}
& \cl{3}{2}[{\bbC(4)\oplus\bbC(4)}]&
\cl{2}{3}[{\bbC(4)\oplus\bbC(4)}]
\\ \hline
1&
t=\tfrac14 (1 + \e{1})(1 + \ii \e{23})
&
t=\tfrac14 (1 + \e{1})(1 + \e{23})
\\
2&
\begin{multlined}[t][0.1\columnwidth]
  \{t 1 t, t \e{45} t \}
\end{multlined}
&
\begin{multlined}[t][0.1\columnwidth]
  \{t 1 t, t \e{45} t \}
\end{multlined}
\\
3&
\begin{multlined}[t][0.1\columnwidth]
\{t, \ii \e{2} t, \ii \e{4} t, \e{24} t \}
\end{multlined}
&
\begin{multlined}[t][0.1\columnwidth]
\{t, \e{2} t, \e{4} t, \e{24} t \}
\end{multlined}
\\
4&
\begin{multlined}[t][0.1\columnwidth]
\{E_{11} - E_{22} - E_{33} + E_{44} + E_{55} - E_{66} - E_{77} + E_{88},\\[-3ex]
  \ii(E_{12} - E_{21} - E_{34} + E_{43} + E_{56} - E_{65} - E_{78} + E_{87}),\\[-1ex]
-E_{12} - E_{21} + E_{34} + E_{43} - E_{56} - E_{65} + E_{78} + E_{87},\\[-1ex]
\ii(-E_{13} - E_{24} - E_{31} - E_{42} + E_{57} + E_{68} + E_{75} + E_{86}),\\[-1ex]
E_{13} + E_{24} - E_{31} - E_{42} + E_{57} + E_{68} - E_{75} - E_{86}
\}
\end{multlined}
&
\begin{multlined}[t][0.1\columnwidth]
\{E_{11} - E_{22} - E_{33} + E_{44} + E_{55} - E_{66} - E_{77} + E_{88},\\[-3ex]
   E_{12} + E_{21} + E_{34} + E_{43} - E_{56} - E_{65} - E_{78} - E_{87},\\[-1ex]
-E_{12} + E_{21} - E_{34} + E_{43}  + E_{56} - E_{65} + E_{78} - E_{87},\\[-1ex]
-E_{13} + E_{24} + E_{31} - E_{42} - E_{57} + E_{68} + E_{75} - E_{86},\\[-1ex]
\ii(-E_{13} + E_{24} - E_{31} + E_{42} + E_{57} - E_{68} + E_{75} - E_{86})
\}
\end{multlined}
\\
5&
\begin{multlined}[t][0.1\columnwidth]
\Psi=(s_{1} + s_{2}\e{45} + \ii s_{3}\e{2} + \ii s_{4}\e{245} + \ii s_{5}\e{4} + \ii s_{6}\e{5}\\[-3ex] + s_{7}\e{24} + s_{8}\e{25}) t
\end{multlined}
&
\begin{multlined}[t][0.1\columnwidth]
\Psi=(s_{1} + s_{2}\e{45} + s_{3}\e{2} + s_{4}\e{245} + s_{5}\e{4} + s_{6}\e{5}\\[-3ex] + s_{7}\e{24} + s_{8}\e{25}) t
\end{multlined}
\\
6&
\begin{multlined}[t][0.1\columnwidth]
 (s_{1} + \ii s_{2})E_{11} + (s_{3} + \ii s_{4})E_{21}   + (s_{5} - \ii s_{6})E_{31} \\[-3ex]+ (s_{7} - \ii s_{8})E_{41} +(s_{1} - \ii s_{2})E_{55} + (s_{3} - \ii s_{4})E_{65} \\[-1ex]+ (-s_{5} - \ii s_{6})E_{75} + (-s_{7} - \ii s_{8})E_{85}
\end{multlined}
&
\begin{multlined}[t][0.1\columnwidth]
 (s_{1} + \ii s_{2})E_{11} + (s_{3} + \ii s_{4})E_{21}  + (s_{5} - \ii s_{6})E_{31} \\[-3ex] + (s_{7} - \ii s_{8})E_{41} +(s_{1} - \ii s_{2})E_{55} + (-s_{3} + \ii s_{4})E_{65} \\[-1ex]+ (s_{5} + \ii s_{6})E_{75} + (-s_{7} - \ii s_{8})E_{85}
\end{multlined}
\\
7&
\begin{multlined}[t][0.1\columnwidth]
-\langle\e{45}\reverse{\Psi}^*\e{45}\Psi\rangle=\tfrac14(|s_1|^2+\cdots+|s_8|^2)
\end{multlined}
&
\begin{multlined}[t][0.1\columnwidth]
  -\langle\e{12}\cliffordconjugate{\Psi}^*\e{12}\Psi\rangle=\tfrac14(|s_1|^2+\cdots+|s_8|^2)
\end{multlined}
\end{tabular}
\vskip 10pt

\begin{tabular}{l|>{$}l<{$}>{$}l<{$}>{$}l<{$}}
& \cl{1}{4}[{\bbC(4)\oplus\bbC(4)}]&
\cl{0}{5}[{\bbC(4)\oplus\bbC(4)}]
\\ \hline
1&
t=\tfrac14 (1 + \e{1})(1 + \ii \e{23}) &
t=\tfrac14 (1 + \ii\e{1})(1 + \ii\e{23})
\\
2&
\begin{multlined}[t][0.1\columnwidth]
  \{t 1 t, t \e{45} t \}
\end{multlined}
&
\begin{multlined}[t][0.1\columnwidth]
  \{t 1 t, t \e{45} t \}
\end{multlined}
\\
3&
\begin{multlined}[t][0.1\columnwidth]
\{t, \ii \e{2} t, \ii \e{4} t, \e{24} t \}
\end{multlined}
&
\begin{multlined}[t][0.1\columnwidth]
\{t, \e{2} t, \ii \e{4} t, \ii \e{24} t\}
\end{multlined}
\\
4&
\begin{multlined}[t][0.1\columnwidth]
\{E_{11} - E_{22} - E_{33} + E_{44} + E_{55} - E_{66} - E_{77} + E_{88},\\[-3ex]
  \ii(-E_{12} - E_{21} + E_{34} + E_{43} - E_{56} - E_{65} + E_{78} + E_{87}),\\[-1ex]
-E_{12} + E_{21} + E_{34} - E_{43} - E_{56} + E_{65} + E_{78} - E_{87},\\[-1ex]
\ii(-E_{13} - E_{24} - E_{31} - E_{42} + E_{57} + E_{68} + E_{75} + E_{86}),\\[-1ex]
E_{13} + E_{24} - E_{31} - E_{42} + E_{57} + E_{68} - E_{75} - E_{86}
\}
\end{multlined}
&
\begin{multlined}[t][0.1\columnwidth]
  \{\ii (-E_{11} + E_{22} + E_{33} - E_{44} - E_{55} + E_{66} + E_{77} - E_{88}),\\[-3ex]
  -E_{12} + E_{21} - E_{34} + E_{43}  + E_{56} - E_{65} + E_{78} - E_{87},\\[-1ex]
  \ii (E_{12} + E_{21} + E_{34} + E_{43} - E_{56} - E_{65} - E_{78} - E_{87}),\\[-1ex]
  \ii (-E_{13} + E_{24} - E_{31} + E_{42}  + E_{57} - E_{68} + E_{75} - E_{86}),\\[-1ex]
E_{13} - E_{24} - E_{31} + E_{42} + E_{57} - E_{68} - E_{75} + E_{86}
\}
\end{multlined}
\\
5&
\begin{multlined}[t][0.1\columnwidth]
\Psi=(s_{1} + s_{2}\e{45}  + \ii s_{3}\e{2} + \ii s_{4}\e{245}  + \ii s_{5}\e{4} + \ii s_{6}\e{5}\\[-3ex] + s_{7}\e{24} + s_{8}\e{25}) t
\end{multlined}
&
\begin{multlined}[t][0.1\columnwidth]
\Psi=(s_{1} + s_{2}\e{45}  + s_{3}\e{2} + s_{4}\e{245}  + \ii s_{5}\e{4} + \ii s_{6}\e{5}\\[-3ex] + \ii s_{7}\e{24} + \ii s_{8}\e{25}) t
\end{multlined}
\\
6&
\begin{multlined}[t][0.1\columnwidth]
 (s_{1} + \ii s_{2})E_{11} + (s_{3} + \ii s_{4})E_{21}  + (s_{5} - \ii s_{6})E_{31}+ (s_{7} \\[-3ex] - \ii s_{8})E_{41} +(s_{1} - \ii s_{2})E_{55} + (s_{3} - \ii s_{4})E_{65} \\[-1ex]+ (-s_{5} - \ii s_{6})E_{75} + (-s_{7} - \ii s_{8})E_{85}
\end{multlined}
&
\begin{multlined}[t][0.1\columnwidth]
 (s_{1} + \ii s_{2})E_{11} + (s_{3} + \ii s_{4})E_{21}   + (s_{5} - \ii s_{6})E_{31}\\[-3ex] + (s_{7} - \ii s_{8})E_{41} +(s_{1} - \ii s_{2})E_{55} + (-s_{3} + \ii s_{4})E_{65} \\[-1ex]+ (-s_{5} - \ii s_{6})E_{75} + (s_{7} + \ii s_{8})E_{85}
\end{multlined}
\\
7&

\langle\e{1}\reverse{\Psi}^*\e{1}\Psi\rangle=\tfrac14(|s_1|^2+\cdots+|s_8|^2)
&
\langle\cliffordconjugate{\Psi}^*\Psi\rangle=\tfrac14(|s_1|^2+\cdots+|s_8|^2)
\end{tabular}
\caption{Complex spinor tables for geometric algebras with vector space dimension $n=5$}\label{sptc5}
\end{table}
\vskip 10pt

\end{flushleft}


\bibliography{algebraic}

\end{document}